\documentclass{jfm}
\usepackage{graphicx}
\usepackage{epstopdf, epsfig}
\usepackage[T1]{fontenc}

\usepackage{amssymb}
\usepackage{amsfonts}
\usepackage{graphicx,epsfig}
\usepackage{amsmath}
\usepackage{float}
\usepackage{color}
\usepackage{subfigure}
\usepackage{natbib}
\usepackage{upgreek}  % Greek letter non italic

\shorttitle{Chemiosomotic flow in a narrow fluidic channel}
\shortauthor{Pranab Kumar Mondal}

\title{Chemiosmotic flow in a narrow fluidic channel}

\author{Pranab Kumar Mondal
\corresp{\email{pranabm@iitg.ac.in}}}
\affiliation{Microfluidics and Microscale Transport Processes Laboratory, Department of Mechanical Engineering, Indian Institute of Technology Guwahati, Guwahati, Assam 781039, India}
\begin{document}
\maketitle

\begin{abstract}

A liquid volume containing dissolved solutes moves through a charged nanofluidic channel under the influence of the concentration gradient of the solutes, non-trivially modulated by the electrostatic interaction between ionic liquid and charged surface. The available studies in this paradigm primarily focus on either of diffusioosmosis or electrodiffusioosmosis modulated physicochemical hydrodynamical phenomenon, essentially to obtain a net throughput at the overlapping scales. Here, we develop a theoretical model that accounts for the induced pressure gradient stemming from the concentration gradient of the solutes alongside the axially varying electrical double layer effect in tandem and characterizes the chemiosmotic flow in a reservoir-connected nanofluidic system. Starting from the potential distribution developed due to the solute gradient modulated electrical double layer effect, we look at the effect of pertinent physicochemical parameters and their eventual manifestations onto the purely chemiosmotic transport, aptly described in this endeavor. We analytically establish a chemiosmotic velocity scale from a macroscopic viewpoint, relating flow velocity with the relevant parameters, and uniquely measuring the magnitude of chemiosmotic velocity. A closer as well as consistent agreement on theoretical predictions with the corresponding full-scale simulated results, both in the limit and beyond the Debye-Huckel approximation, substantiates the efficacy of our theory.

\end{abstract}

\begin{keywords}
Chemiosmosis; Nanochannel; Axially varying electric double layer (EDL); Velocity scale; Lubrication approximation
\end{keywords}

\section{Introduction}

Spontaneous migration of small fluid volume through a narrow fluidic channel has been of prime interest to multiple communities, attributed primarily to the wide gamut of practical applications of this fluidic operation in different areas of science and engineering \citep{stone2004engineering,kirby2010micro}. In narrow fluidic systems/devices, different mechanisms/flow actuation parameters such as applied pressure gradient, electroosmosis, thermocapillarity, Magnetofluidics, acoustofluidics, and optical actuation have been employed to drive liquid in the configured pathway quite conveniently \citep{mondal2013electric,dasgupta2014thermocapillary,mondal2014pulsating,shyam2020dynamics,shyam2021magnetofluidic,friend2011microscale}. It is worth mentioning here that liquid manipulation in small scale systems involving gradient of solute concentration seems to be attractive and have gained practical relevance as well because of the inherent non-invasiveness of this driving mechanisms \citep{ault2018diffusiophoresis,keh2016diffusiophoresis,shim2022diffusiophoresis,anderson1989colloid}. To this end, researchers have explored several physico-chemical hydrodynamical phenomena, such as diffusio-osmosis, diffusiophoresis, electrodiffusiosmosis, chemiphoresis, and established the dynamical characteristics of the underlying electrolyte transport associated to these flow actuation processes at the prevailing spatio-temporal scales \citep{keh2016diffusiophoresis,shim2022diffusiophoresis,qian2007diffusioosmotic,karimzadeh2022blue}. As reported in the referred analyses \citep{keh2009diffusioosmotic,keh2005diffusioosmosis}, the resulting flow dynamics pertaining to the flow configuration mentioned above is strongly governed by the cooperative-correlative effects of fluid mechanics, solute diffusion, and the interfacial electrostatics. Despite the mentioned flow actuation parameters being aptly studied for the electrolytic solute transport in different fluidic configurations, the potential aspect of chemiosmosis or chemiosmotic effect in solute (small electrolyte volume) pumping by interconnecting physico-chemical hydrodynamics at the overlapping scales, is at variance \citep{hanson1978application,hsu2014stationary,pandey2023chemiosmotic}.

In the paradigm of low Reynolds number transport, the intervening effect of viscous force and the prevailing driving force due to the concentration gradient of the constituent solutes in the presence of charged solid wall has an important role in most of the flow actuation mechanisms discussed above \citep{palacci2010colloidal,palacci2012osmotic}. In this context, application of an external electric field, which is typical to diffusio-osmosis, diffusiophoresis, electrodiffusio-osmosis, alters the forcing being applied over the fluid mass in the liquid-solid interfacial region, and simultaneously tunes the gradients of electrolytic solute under motion in the fluidic  pathway. The spontaneous motion of ionic liquid in the pathway that sets in and occurs due to the concentration gradient of the solutes, also referred to as the chemiosmosis \citep{keh2016diffusiophoresis,keh2009diffusioosmotic,keh2005diffusioosmosis}, is governed by the imposed concentration gradient modulated electrical double layer effect as well. However, the analysis available in the literature on this part is either focused on the diffusio-osmosis \citep{ajdari2006giant,shim2022diffusiophoresis,probstein2005physicochemical} or electrodiffusio-osmosis \citep{ismayeel2023prediction} modulated physicochemical hydrodynamical phenomenon. It is worth adding here that these two flow actuation mechanisms are involved with the simultaneous effects of applied solute gradient and electric field while bringing the liquid in motion through the pathway \citep{anderson1982motion,prieve1984motion}.

Pertaining to the transport of ionic solution involving the applied concentration gradient of solute only, interconnections of induced pressure gradient due to the prevailing solute concentration gradient and the forcing that stems from the axially varying potential gradient inside the EDL provide a suitable means of manipulating the underlying flow through the narrow fluidic channel. Albeit such a configuration is deemed pertinent to reveal the basic chemiosmotic effect on the underlying transport at the nanofluidic scale, this aspect has not been looked at by the researchers' to date in the open literature. Here, we discuss chemiosmotic transport in detail from the perspectives of theoretical analysis and numerical simulations. The characteristics of the axial chemiosmotic velocity have been thoroughly investigated with respect to the fundamental electrokinetic parameters that primarily generate the flow itself. In this endeavor, under the Debye-H\"uckel limit, we establish a chemiosmotic slip velocity scale analytically for thin EDL scenarios. Such a detailed analysis of the characteristics of chemiosmotic flow within a uniformly charged nanochannel, as discussed in this article, seems to provide a detailed understanding of this phenomenon, which has not been properly addressed to date as far as the author's knowledge.

\section{Mathematical modeling and perturbed solution}
\begin{figure}
  \centering
  \subfigure{\includegraphics[width=4in]{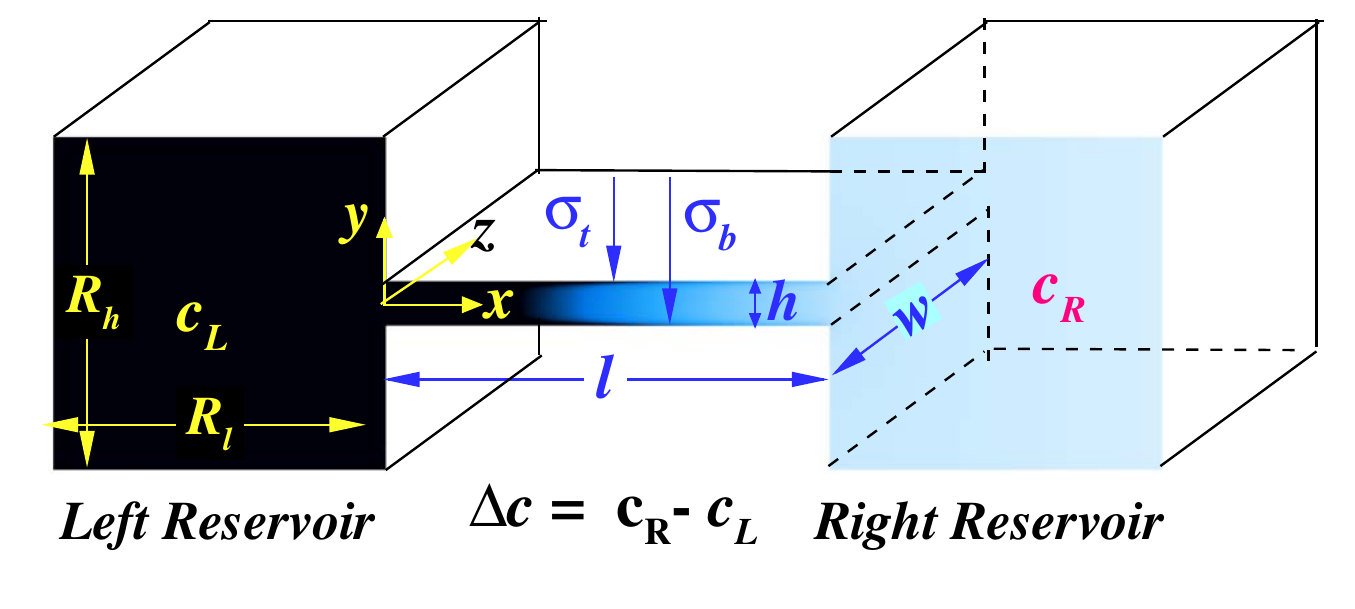}}
  \caption{Schematic of a nanochannel of height $h$, length $l$ which connects two identical reservoirs of dimension $R_h\times R_l$ on both sides of the nanochannel ends. The width of the nanochannel and the reservoirs in $z-$direction is sufficiently higher and thus neglected in the present study ($h\ll w$). The top and bottom walls of the nanochannel bear uniform surface charge density $\sigma_t$ and $\sigma_b$, respectively. The nanochannel is considered under a fixed concentration difference $\Updelta c$ along the opposite axial direction by imposing a higher (lower) bulk concentration $c_L$ ($c_R$) in the left (right) side reservoir.} 
  \label{fig_model}
\end{figure}

As presented in figure~\ref{fig_model}, we consider a nanochannel of height $h$, length $l$, connecting two identical reservoirs of size $R_h\times R_l$ at both ends. The width $w$ of the nanochannel is considered to be sufficiently large compared to the nanochannel height, i.e., $h\ll w$. This consideration allows us to study the present problem using a two-dimensional ($x,y$) coordinate system. The coordinate system is attached to the left center of the channel, as shown in figure~\ref{fig_model}. The nanochannel and reservoirs are filled with an incompressible, Newtonian, binary, monovalent electrolyte with a fixed density $\rho$, permittivity $\epsilon_r\epsilon_0$, and viscosity $\mu$ at room temperature $T$. In order to create a constant concentration difference $\Updelta c$ along the nanochannel axis, the left and right reservoirs are maintained at a fixed bulk concentration $c_L$ and $c_R$. Without any loss of generality, a negative concentration gradient $(\Updelta c/l<0)$ along the axial direction is maintained by considering a higher bulk concentration in the left reservoir, i.e., $c_L\geq c_R$. Accounting for the prevailing negative concentration gradient across the nanochannel, the highly concentrated electrolyte tends to approach a low concentration regime following the natural diffusion process. The flow occurs along the axial direction in accordance with the pressure difference generated due to the solute concentration gradient. Since the nanochannel walls possess some uniform surface charge density, the counterions are attracted towards the charged wall and generate a charged cloud, better known as the electric double layer (EDL) \citep{masliyah2006electrokinetic}. The thickness of this EDL is characterized by $\kappa^{-1}=\sqrt{\epsilon_0\epsilon_r k_BT/(2ZeFc_\infty)}$, where $\epsilon_0$ represents the vacuum permittivity, $k_B$ stands for Boltzmann constant, $F$ denotes the Faraday constant, and $c_\infty(x,0)$ [mM] depicts the local bulk molar concentration along the central line. This indicates that the local bulk concentration and, thus, the local thickness of the EDL varies spatially. For the sake of obtaining a constant EDL thickness, we refer to the absolute reference concentration as $c_0=(c_L+c_R)/2$, which leads us to derive a fixed dimensionless Debye parameter $\kappa_0 h$, with $\kappa^{-1}_0=\sqrt{\epsilon_0\epsilon_r k_BT/(2ZeFc_0)}$. Since the local EDL thickness spatially varies, it generates a non-vanishing axial electric field across the nanochannel. The surplus mobile ions in the EDL start to move in the direction of the axial electric field and generate a flow in the axial direction. 

We next take an effort to obtain the closed-form expression of the electric double layer potential, the pressure gradient, the non-zero axial electric field, and the corresponding axial velocity profiles using an analytical framework consistent with the lubrication approximation ( $h\ll l$). In doing so, we consider sufficiently long nanochannel and appeal to the Debye-H\"uckel approximation. The solution procedure is aptly outlined in the forthcoming Sections. 

\subsection{Ion distribution and Electric potential}
The distribution and transport phenomena of ionic concentration of $k^{th}$ ionic species for a dilute electrolyte solution (below $10^3$ mM \citep{levine1975theory}) can be described by the Nernst-Planck equations as given below in Eq. (\ref{NP_equ_1}). Note that to obtain this equation (\ref{NP_equ_1}), we neglect the finite ion size effects and interionic correlations, while a sufficiently small electric field is considered as well. 
\begin{equation}
\frac{\partial c_k}{\partial t}+\nabla\cdot\Big(\mathbf{u}c_k-D_k\nabla c_k-D_kc_k\frac{z_kF}{RT}\nabla\phi\Big)=0.
\label{NP_equ_1}
\end{equation}
Here, the term within the bracket denotes the net flux, which consists of a linear superimposition of advection $(\mathbf{u}c_k)$, diffusion $(-D_k\nabla c_k)$, and electromigration $(-z_k D_k c_k F \nabla \phi/RT)$ fluxes under the framework of linear non-equilibrium thermodynamics. Using suitable reference scales, we write the ion-conservation equation in its dimensionless form. In doing so, we normalize the axial $x-$coordinate by nanochannel length $l$, and lateral $y-$coordinate by nanochannel height $h$, i.e., $(X,Y)=(x/l,y/h)$. In order to normalize the velocity field $\mathbf{u}(=(u,v))$, we define $(U,V)=(u/U_0,v/V_0)$ where $V_0=\delta U_0$ and $\delta=h/l$. The electric potential $(\phi)$ and ionic concentration $(c_k)$ are scaled by thermal potential $(k_BT/Ze)$ and reference concentration $(c_0)$, respectively, i.e., $\Phi=\phi/(k_BT/Ze)$ and $C_k=c_k/c_0$. With $\tau=tU_0/l$ as the normalized time, and $Pe_k=U_0h/D_k$ being the ionic P\'eclet number, the normalized Nernst-Planck equations take the following form
\begin{equation}
\begin{split}
Pe_k \delta \frac{\partial C_k}{\partial \tau} - \Bigg[  \delta^2 \frac{\partial^2 C_k}{\partial X^2} + \frac{\partial^2 C_k}{\partial Y^2} 
 + \delta^2 Z_k \frac{\partial}{\partial X}\left( C_k\frac{\partial \Phi}{\partial X}\right) + Z_k \frac{\partial}{\partial Y}\left( C_k\frac{\partial \Phi}{\partial Y}\right) \\
 - Pe_k \delta \frac{\partial}{\partial X}\left( C_k U\right) - Pe_k \delta \frac{\partial}{\partial Y}\left( C_k V\right)\Bigg]=0.
 \end{split}
\label{NP_equ_2}
\end{equation}

It is commonly considered in electrokinetics that the total electric potential $\phi(x,y)(=\psi(x,y)+\varphi)$. As seen, total electric potential consists of the linear superposition of the electric double layer potential $\psi(x,y)$ at the electrolyte-solid interface referring to the deviation from electroneutrality \citep{fair1971reverse}, and induced electric potential $\varphi$. Being associated with vanishing charge density, the Laplacian operator acting on the induced potential $\varphi$ disappears. Pertinent to the symmetry condition along the central line, we have $\varphi=\varphi(x)$ and $d_x\varphi=-E$ is a constant. The independence of $\varphi$ with respect to $y-$coordinate allows $\partial_y\phi=\partial_y\psi$ or equivalently $\partial_Y\Phi=\partial_Y\Psi$ in normalized form, where $\partial_Y\equiv \partial/\partial Y$.

Considering the applicability of the lubrication approximation, we have the nanochannel height sufficiently smaller than its length in this endeavor, i.e., $h\ll l$ or $\delta\ll1$. Also, for simple electrolytes, typically considered in nanofluidic applications, the ionic P\'eclet number $Pe_k<1$ \citep{masliyah2006electrokinetic}. Thus, neglecting the terms multiplied with $\delta^2$, $Pe_k\delta$, and considering $\partial_Y\Phi=\partial_Y\Psi$, one finds from Eq.(\ref{NP_equ_2}),
$ \partial_Y[ \partial_Y C_k+z_kC_k\partial_Y\Psi] =0$. Integrating this equation leads us to obtain the following,
\begin{equation}
    \frac{\partial C_k}{\partial Y}+ Z_kC_k\frac{\partial\Psi}{\partial Y}=f(X),
    \label{NP_equ_3}
\end{equation}
where $f(X)$ is a function of $X$ comes as an integration constant. However, the symmetry conditions along the central line ($Y=0$) allow us to consider $f(X)$ to be identically equal to zero. Another integration of Eq.(\ref{NP_equ_3}) gives us an ionic distribution similar to the Boltzmann distribution as written below.
\begin{equation}
    C_k(X,Y)=C_k^{\infty}(X)e^{-Z_k\Psi(X,Y)}.
    \label{NP_equ_4}
\end{equation}
Here, $C_k^\infty(=c_k^\infty/c_0)$ is the normalized bulk ionic concentration of $k^{th}$ species along the central line of the nanochannel. The definition of $\Psi$ gives the essence of deviation from the electroneutrality \citep{fair1971reverse}. Thus, $C_k^\infty$ must correspond to the scenario under vanishing charge density. Pertaining to the present study, the consideration of binary symmetric electrolyte implies identical $C_k^\infty$ for both ionic species, i.e., $C_k^\infty(X)=C^\infty(X)$ for $k=1,2$.

The electric potential $\phi$ follows the Poisson equation
\begin{equation}
    \nabla\cdot(\epsilon_r\epsilon_0\nabla\phi)=-\rho_f,
    \label{P_equ_1}
\end{equation}
where the $\epsilon_0\epsilon_r$ represents the permittivity of the electrolyte and $\rho_f(=F\sum_k z_kc_k)$, $k=1,2$ denotes the space charge density. Considering $(\Phi,\Psi)=(\phi,\phi)/(k_BT/Ze)$, $\Phi(X,Y)=\Psi(X,Y)+\mathbf{\varphi}(X)$ and by making use of the normalized Poisson equation (\ref{P_equ_1}), we have
\begin{equation}
\delta^2\frac{\partial }{\partial X^2}(\Psi+\varphi) + \frac{\partial}{\partial Y^2}(\Psi)=-(\kappa_0h)^2\sum_{k=1}^2z_kC_k
    \label{P_equ_2}
\end{equation}
Neglecting the first term in Eq.(\ref{P_equ_2}) as $\delta^2\ll1$, and by substituting Eq.(\ref{NP_equ_4}) in Eq.(\ref{P_equ_2}) one can derive
\begin{equation}
\frac{\partial^2\Psi}{\partial Y^2}=(\kappa h)^2\sinh{\Psi}
    \label{P_equ_3}
\end{equation}
where $\kappa h=\kappa h(X)$ is the local Debye parameter, which depends upon the axial coordinate, and it is related to the global Debye parameter as 
\begin{equation}
\kappa=\kappa_0\sqrt{C^\infty(X)}
    \label{P_equ_4}
\end{equation}
We consider the top and bottom walls of the nanochannel to bear a uniform surface charge density $\sigma_t$ and $\sigma_b$, respectively. This consideration results in Neumann boundary conditions as $\partial_Y\Psi=\sigma_T$ and $\partial_Y\Psi=-\sigma_B$ at $Y=\pm1/2$, respectively. Here, the normalized surface charge densities are derived as $\sigma_{(T/B)}=\sigma_{(t/b)}Zeh/\epsilon_0\epsilon_r k_BT$. The Poisson Eq.(\ref{P_equ_3}) for chemiosmotic flow resembles the classical Poisson equation in electrokinetics, except the only difference is $\kappa=(\kappa(X)$ is a spatial function. However, Eq.(\ref{P_equ_3}) can be solved analytically with a suitable combination of boundary conditions in a similar fashion as is usually employed for a purely electroosmotic flow. In accordance with the surface charge boundary conditions mentioned above, one can derive the electric potential as (in weak form with respect to axial coordinate)
\begin{equation}
\Psi(X,Y)=\frac{1}{2}\left[ \frac{\sigma_T-\sigma_B}{\kappa h\cosh{(\kappa h/2)}}  \sinh{(\kappa hY)}  +  \frac{\sigma_T+\sigma_B}{\kappa h\sinh{(\kappa h/2)}}  \cosh{(\kappa hY)}  \right].
\label{P_equ_5}
\end{equation}
Here, the electric potential $\Psi$ changes with respect to the axial coordinate because the electric double layer thickness $(\kappa h(X))$ also changes along the nanochannel. This axial variation in the EDL potential $\Psi$ gives rise to a non-zero axial electric field which can be obtained as $-\partial_X\Psi=\partial_{\Bar{\kappa}}\Psi d_X \Bar{\kappa}$ with $\Bar{\kappa}=\kappa h$. The expression of $\partial_X\Psi$ reads as,
\begin{equation}
\begin{split}
\partial_X\Psi=\frac{ d_X C^\infty (\sigma_T-\sigma_B)}{ 8(\kappa h)^2\sqrt{C^\infty(X)}}\Bigg[ \frac{2\kappa hY\cosh{(\kappa h Y)} -\sinh{(\kappa hY)}\big[2+\kappa h\tanh{(\kappa h/2)}\big] }{\cosh{(\kappa h/2)}}\Bigg] \\
+ \frac{ d_X C^\infty (\sigma_T+\sigma_B)}{ 8(\kappa h)^2\sqrt{C^\infty(X)}}\Bigg[ \frac{2\kappa hY\sinh{(\kappa h Y)} -\cosh{(\kappa hY)}\big[2+\kappa h\coth{(\kappa h/2)}\big] }{\sinh{(\kappa h/2)}}\Bigg].
\end{split}
\label{P_equ_6}
\end{equation}
In the present study, the concentration gradient (dimensional or dimensionless) is considered as constant, i.e., $d_XC^\infty$ is constant.

\subsection{Momentum transport}
The chemiosmotic flow (COF) field $\mathbf{u}$ within the uniformly charged nanochannel filled with an incompressible, homogeneous electrolyte having a constant density and viscosity, is governed by the Navier-Stokes equation, as given in Eq.(\ref{NS_eq_1}). Note that the last term of Eq.(\ref{NS_eq_1}) is the electrical body force that originates from the Maxwell stress $\epsilon_0\epsilon_r[\nabla\phi\nabla\phi-\nabla\phi\cdot\nabla\phi\mathbf{I}/2]$. Considering the nanochannel is aligned in such a way that gravity acts in a perpendicular direction, thus neglecting the buoyancy effects, the equation can be expressed as
\begin{equation}
\rho\bigg[\frac{\partial \mathbf{u}}{\partial t} + \mathbf{u}\cdot\nabla \mathbf{u}\bigg]=-\nabla p + \nabla\cdot\bigg[ \mu  \bigg\{\nabla\mathbf{u} +\Big(\nabla\mathbf{u}\Big)^T \bigg\}\bigg] - \rho_f\nabla\phi.
  \label{NS_eq_1}
\end{equation}
Substituting the value of space charge density $\rho_f(=F\sum_kz_kc_k)$, electric potential $\phi(=\psi+\varphi)$, and normalize the dependent variables such as the velocity $\mathbf{u}(=(u,v))$ as $(U,V)=(u/U_0,v/\delta U_0)$, pressure as $p/(\mu U_0/h\delta)=P$, electric potential by thermal potential ($\phi/(k_BT/Ze)=\Phi)$, the $X-$momentum equation can be obtained as
\begin{equation}
\begin{split}
\delta Re \Bigg[ \frac{\partial U}{\partial \tau}+ U\frac{\partial U}{\partial X} + V\frac{\partial U}{\partial Y}\Bigg]=- \frac{\partial P}{\partial X} + \Bigg[ \delta^2\frac{\partial^2 U}{\partial X^2}+\frac{\partial^2 U}{\partial Y^2}\Bigg]
    -\frac{Ha}{2}\big(C_+-C_-\big)\Bigg(\frac{\partial\Psi}{\partial X} -\Bar{E}\Bigg),  
    \end{split}
    \label{NS_eq_2}
\end{equation}
and the $Y-$momentum equation can be written as
\begin{equation}
%\begin{split}
    \delta^3 Re \Bigg[ \frac{\partial V}{\partial \tau}+ U\frac{\partial V}{\partial X} + V\frac{\partial V}{\partial Y}\Bigg]=- \frac{\partial P}{\partial Y} + \delta^2\Bigg[ \delta^2\frac{\partial^2 V}{\partial X^2}+\frac{\partial^2 V}{\partial Y^2}\Bigg]
    -\frac{Ha}{2}\big(C_+-C_-\big)\frac{\partial\Psi}{\partial Y}.
 %   \end{split}
 \label{NS_eq_3}
\end{equation}
The normalized momentum equations have two dimensionless parameters, i.e., the Reynolds number $Re=\rho U_0h/\mu$ and Hartmann number $Ha=2\delta hc_0N_Ak_BT/U_0\mu$, where $N_A$ stands for the Avogadro number. In the present study, the flow is highly laminar ($Re\ll1$), and the scaled Hartmann number represents the velocity ratio due to osmotic pressure with the reference velocity. Neglecting the terms of $\mathcal{O}(\delta^2)$ or higher in the $Y-$momentum Eq.~(\ref{NS_eq_3}), and considering $\partial_Y C_\pm=\mp C_\pm\partial_Y\Psi$ with $\partial_Y\equiv \partial/\partial Y$, from  Eq.~(\ref{NP_equ_3}) one can derive the following
\begin{equation}
    \frac{\partial}{\partial Y}\Big[ P - \frac{Ha}{2}(C_++C_-)\Big]=0.
    \label{NS_eq_4}
\end{equation}
Integrating the above Eq.(\ref{NS_eq_4}) with respect to the transverse coordinate, one gets $P-(Ha/2)(C_++C_-)=g(X)$, where $g(X)$ serves as the integration constant. The value of this constant at bulk is equal to the osmotic pressure at bulk, i.e., $[p_0-k_BTN_A\sum_kc_k^\infty]/(\mu U_0/h\delta)=[P_0(X)-HaC^\infty(X)]$ in the scaled form \citep{masliyah2006electrokinetic}. Substituting Eq.~(\ref{NP_equ_4}), the normalized pressure can be expressed as
\begin{equation}
    P(X,Y)=P_0(X)+HaC^\infty(X)\Big[ \cosh{(\Psi(X,Y))}-1\Big],
    \label{NS_eq_5}
\end{equation}
where $P_0(X)$ is the externally applied pressure along the axial direction. Differentiating Eq.(\ref{NS_eq_5}) partially with respect to axial coordinate, one can derive the following
\begin{equation}
    \frac{\partial P}{\partial X}=\frac{d P_0}{dX} + Had_XC^\infty\Big[ \cosh{(\Psi(X,Y))}-1\Big] + HaC^\infty \sinh{(\Psi)} \frac{\partial\Psi}{\partial X},
    \label{NS_eq_6}
\end{equation}
where $d_XC^\infty$ is the concentration gradient along the axial direction and is constant in the present study. Neglecting terms with $\mathcal{O}(\delta^2)$, $Re\delta$, and substituting the axial pressure gradient from Eq.(\ref{NS_eq_6}) in the $X-$momentum Eq.(\ref{NS_eq_2}), the same can be arranged in the form 
\begin{equation}
  \frac{\partial^2U}{\partial Y^2}=\frac{d P_0}{dX} + Ha\Bar{E}C^\infty\sinh{(\Psi)} + Ha d_XC^\infty\big[ \cosh{(\Psi)}-1\big].
  \label{NS_eq_7}
\end{equation}
Under the framework of the Debye-H\"uckel approximation, $\sinh{(\Psi)}\approx\Psi$ and $[\cosh{(\Psi)}-1]\approx\Psi^2/2$ that simplify the Eq.(\ref{NS_eq_7}) further. The simplified equation can be integrated twice with respect to $Y-$coordinate as
\begin{equation}
U=\frac{1}{2}\frac{dP_0}{dX}Y^2+ HaC^\infty\Bar{E}\int\int\Psi dYdY+  \frac{Ha}{2}d_XC^\infty\int\int\Psi^2dYdY+Yf_1(X) + f_2(X),
    \label{NS_eu_8}
\end{equation}
where EDL potential $\Psi(X,Y)$ given in Eq.(\ref{P_equ_5}). Here, $f_1,~f_2$ are integration constants, which can be obtained from the no-slip boundary conditions employed at nanochannel walls, i.e., $U=0$ at $Y=\pm1/2$. After performing some intermediate steps (not given here for conciseness in the presentation. However, the author would like to confirm the availability of this part upon the request of the readers), we obtain the axial velocity profile under Debye-H\"uckel approximation. The profile can be written as
\begin{equation}
\begin{split}
U=-\frac{1}{8}\Big[1-4Y^2\Big]\frac{dP_0}{dX} + \frac{Ha}{2}C^\infty \Bar{E}\Big[  2\omega -(1-2Y)\omega_{\frac{1}{2}} - (1-2Y)\omega_{-\frac{1}{2}}\Big] \\
+ \frac{Ha}{4}d_XC^\infty\Big[  2\Omega -(1-2Y)\Omega_{\frac{1}{2}} - (1-2Y)\Omega_{-\frac{1}{2}}\Big],
\end{split}
    \label{NS_eq_9}
\end{equation}
where $\Omega=\int\int\Psi^2 d^2Y$ and $\omega=\int\int\Psi d^2Y$, with $\Omega_{\pm\frac{1}{2}}=\Omega|_{Y=\pm\frac{1}{2}}$ and $\omega_{\pm\frac{1}{2}}=\omega|_{Y=\pm\frac{1}{2}}$. The expression of these quantities is given below.
\begin{align}
\label{NS_eq_10}
\begin{split}
~~\Omega &=\frac{(\sigma_T-\sigma_B)^2}{4(\kappa h)^2 \cosh^2{(\kappa h/2)}}\bigg[-\frac{Y^2}{4} +\frac{ \cosh{(2\kappa hY)} }{8(\kappa h)^2} \bigg]  +   \frac{(\sigma_T+\sigma_B)^2}{4(\kappa h)^2 \sinh^2{(\kappa h/2)}}\bigg[\frac{Y^2}{4}  +\frac{ \cosh{(2\kappa hY)} }{8(\kappa h)^2} \bigg]  \\
   & + \frac{(\sigma_T^2-\sigma_B^2)}{2(\kappa h)^2 \sinh{(\kappa h/2)}}\bigg[\frac{ \sinh{(2\kappa hY)} }{4(\kappa h)^2} \bigg] 
\\
~\Omega_{\frac{1}{2}} & =\frac{(\sigma_T-\sigma_B)^2}{4(\kappa h)^2 \cosh^2{(\kappa h/2)}}\bigg[-\frac{1}{16} +\frac{ \cosh{(\kappa h)} }{8(\kappa h)^2} \bigg]  +   \frac{(\sigma_T+\sigma_B)^2}{4(\kappa h)^2 \sinh^2{(\kappa h/2)}}\bigg[\frac{1}{16} +\frac{ \cosh{(\kappa h)} }{8(\kappa h)^2} \bigg]  \\
   & + \frac{(\sigma_T^2-\sigma_B^2)}{2(\kappa h)^2 \sinh{(\kappa h/2)}}\bigg[\frac{ \sinh{(\kappa h)} }{4(\kappa h)^2} \bigg]  
\\
\Omega_{-\frac{1}{2}} & =\frac{(\sigma_T-\sigma_B)^2}{4(\kappa h)^2 \cosh^2{(\kappa h/2)}}\bigg[-\frac{1}{16} +\frac{ \cosh{(\kappa h)} }{8(\kappa h)^2} \bigg]  +   \frac{(\sigma_T+\sigma_B)^2}{4(\kappa h)^2 \sinh^2{(\kappa h/2)}}\bigg[\frac{1}{16} +\frac{ \cosh{(\kappa h)} }{8(\kappa h)^2} \bigg]  \\
& - \frac{(\sigma_T^2-\sigma_B^2)}{2(\kappa h)^2 \sinh{(\kappa h/2)}}\bigg[\frac{ \sinh{(\kappa h)} }{4(\kappa h)^2} \bigg]  
\\
\omega & =\frac{(\sigma_T+\sigma_B)}{2(\kappa h)^3}\bigg[ \frac{\cosh{(\kappa hY)}}{\sinh{(\kappa h/2)}} \bigg] + \frac{(\sigma_T-\sigma_B)}{2(\kappa h)^3}\bigg[ \frac{\sinh{(\kappa hY)}}{\cosh{(\kappa h/2)}} \bigg]
\\
\omega_{\frac{1}{2}} & =\frac{(\sigma_T+\sigma_B)}{2(\kappa h)^3}\bigg[ \frac{1}{\tanh{(\kappa h/2)}} \bigg] + \frac{(\sigma_T-\sigma_B)}{2(\kappa h)^3}\bigg[\tanh{(\kappa h/2)} \bigg]
\\
\omega_{-\frac{1}{2}} & =\frac{(\sigma_T+\sigma_B)}{2(\kappa h)^3}\bigg[ \frac{1}{\tanh{(\kappa h/2)}} \bigg] - \frac{(\sigma_T-\sigma_B)}{2(\kappa h)^3}\bigg[\tanh{(\kappa h/2)} \bigg].
\end{split}
\end{align}
Note that the first two terms of the axial chemiosmotic velocity Eq.(\ref{NS_eq_9}) are due to the applied external pressure gradient and applied external electric field, respectively. These terms vanish in a chemiosmotic flow that sets in solely due to the solute concentration gradient. It is worth adding here that from the perspective of practical use of the proposed fluidic configuration, it is more feasible to measure the net throughput or the net volume flux across the nanochannel. To obtain the chemiosmotic flux analytically under the Debye-H\"uckel approximation, one needs to integrate the axial velocity Eq.(\ref{NS_eq_9}) as $Q=w\int_{-h/2}^{h/2}udy$. Here, $Q$ represents the dimensional volume flux, $w$ is the width of the nanochannel in $z-$direction, and $u(=UU_0)$ is the dimensional axial velocity under Debye-H\"uckel approximation. After integrating, the chemiosmotic volume flux can be expressed as
\begin{align}
\begin{split}
\frac{Q}{whU_0}&=-\frac{1}{12}\frac{dP_0}{dX} + Ha C^\infty \Bar{E}\frac{(\sigma_T+\sigma_B)}{(\kappa h)^3}\bigg[ \kappa h- \frac{1}{2\tanh{(\kappa h/2)}}\bigg] + Ha\frac{d_XC^\infty}{4(\kappa h)^2} \\
& \times\Bigg\{\frac{ (\sigma_T-\sigma_B)^2}{\cosh^2{(\kappa h/2)}}\bigg[-\frac{1}{96} +\frac{ \sinh{(\kappa h)} }{16(\kappa h)^3} \bigg]  +\frac{ (\sigma_T+\sigma_B)^2}{ \sinh^2{(\kappa h/2)}}\bigg[\frac{1}{96} +\frac{ \sinh{(\kappa h)} }{16(\kappa h)^3} \bigg]\Bigg\} \\
&- Ha\frac{d_XC^\infty}{8(\kappa h)^2} \Bigg\{ \frac{(\sigma_T-\sigma_B)^2}{\cosh^2{(\kappa h/2)}}\bigg[-\frac{1}{16} +\frac{ \cosh{(\kappa h)} }{8(\kappa h)^2} \bigg]  +   \frac{(\sigma_T+\sigma_B)^2}{ \sinh^2{(\kappa h/2)}}\bigg[\frac{1}{16} +\frac{ \cosh{(\kappa h)} }{8(\kappa h)^2} \bigg] 
\Bigg\}.
\end{split}
    \label{NS_eq_11}
\end{align}
The first two terms in the RHS of the expression of volume flux are associated with the external pressure gradient and applied electric field, respectively. Certainly, these two terms will disappear in a purely chemiosmotic flow. The last two terms, multiplied with concentration gradient $d_XC^\infty$, correspond to chemiosmotic flow across the nanochannel. The concentration gradient remains constant in the present study.

\section{Numerical framework}
\begin{figure}
  \centering
  \subfigure{\includegraphics[width=2.5in]{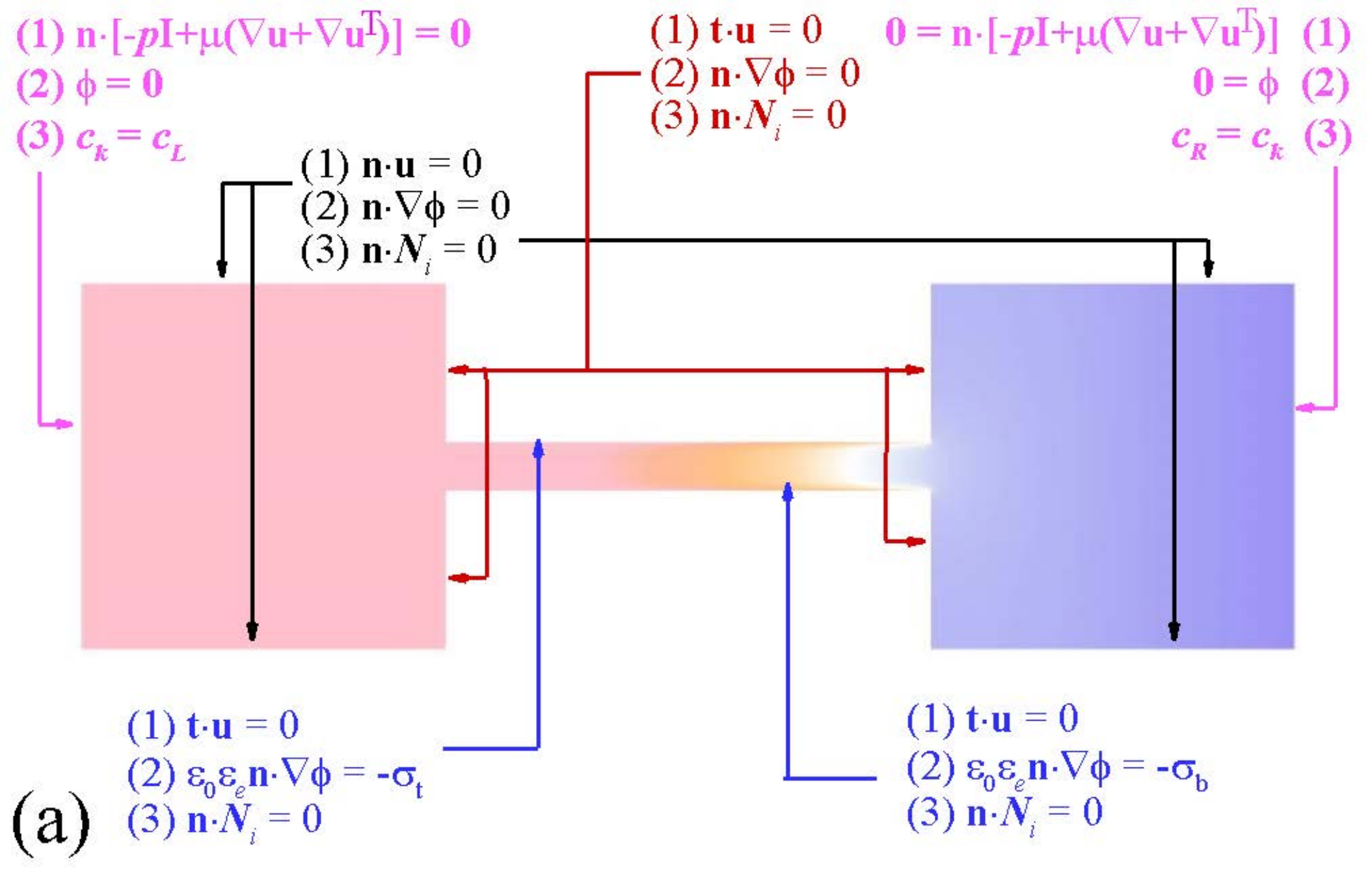}}
  \subfigure{\includegraphics[width=2.5in]{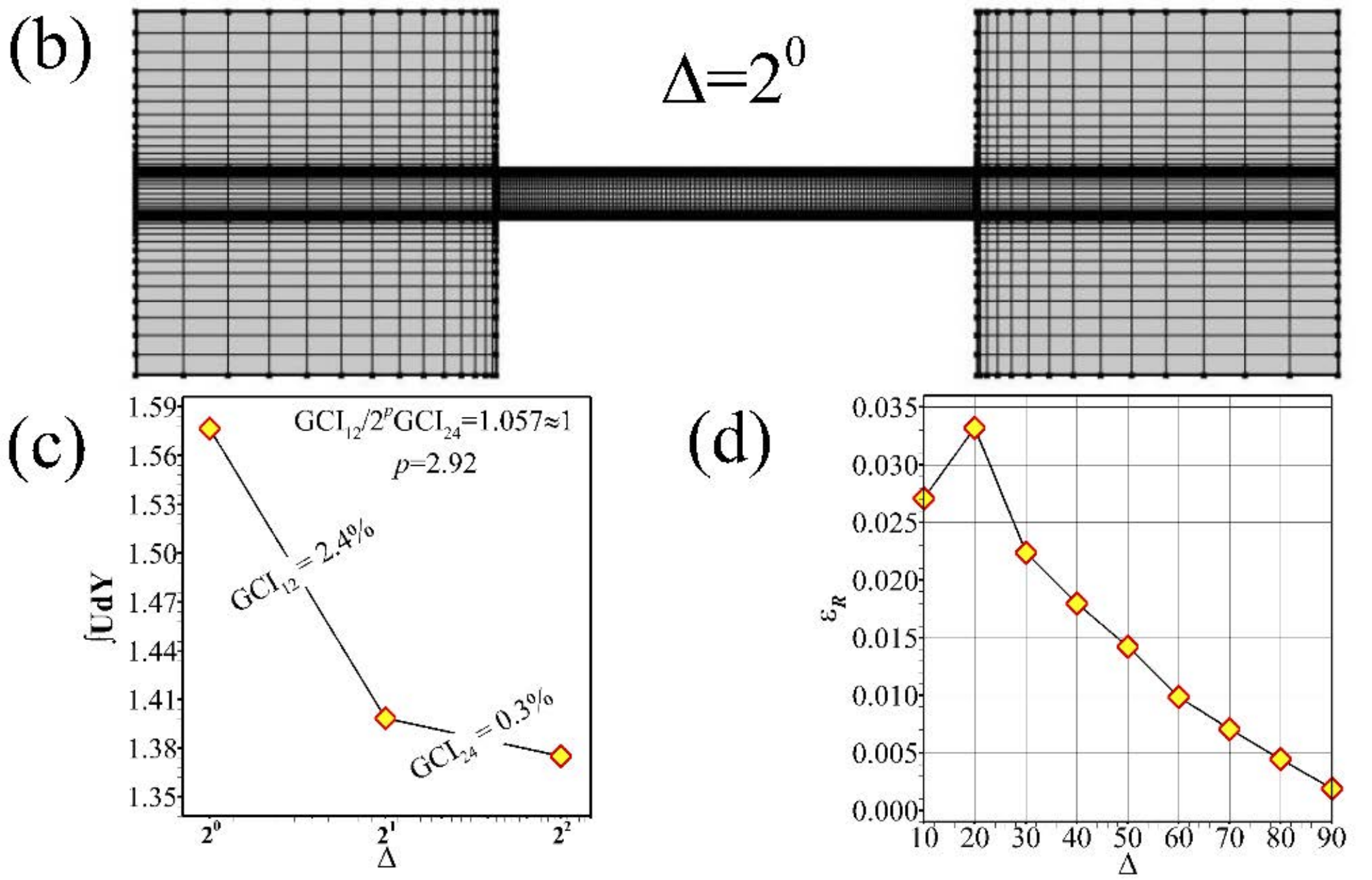}}
  \caption{(a) Boundary conditions (Dimensional) for the electric potential, ionic concentrations, and velocity field at specific boundaries. (b) Grid distribution for $\Updelta=2^0$ (coarser distribution is shown here due to visuality issue), $l=20h$, $R_h=R_l=15h$. (c) Grid independence test is shown for $\Updelta=2^0,2^1,2^2$. (d) Finite-size effect of reservoirs is verified in terms of the relative error $\varepsilon_R$ for several dimensions of the reservoirs. } 
  \label{fig_bc}
\end{figure}

In this endeavor, we also make an effort to model the chemiosmotic flow within the uniformly charged nanochannel and simulate the underlying physicochemical hydrodynamical phenomenon using the framework of COMSOL Multiphysics\,\textsuperscript{\tiny\textregistered}. It is worth mentioning here that the developed numerical model works beyond Debye-H\"uckel approximation and lubrication approximation. We solve the following dimensionless governing equations to model the chemiosmotic flow in charged nanochannel under a steady state.
\begin{align}
\label{num_eq}
\begin{split}
Re[\mathcal{U}\cdot\overline{\nabla}\mathcal{U}] &=-\overline{\nabla}\mathcal{P} +\overline{\nabla}^2\mathcal{U} + (\kappa_0 h)^2(\widetilde{C}_+-\widetilde{C}_-)(\overline{\nabla}\Phi)/2 \\
\overline{\nabla}\cdot\mathcal{U} &=0 \\
\overline{\nabla}^2\Phi & =- (\kappa_0 h)^2(\widetilde{C}_+-\widetilde{C}_-)/2 \\
0 &=  \overline{\nabla}\cdot[ -\overline{\nabla}\widetilde{C}_k - Z_k\widetilde{C}_k\overline{\nabla}\Phi+ Pe_k \widetilde{C}_k\mathcal{U} ]
\end{split}
\end{align}
Here, all the spatial coordinates are scaled by channel height $h$, i.e., $\overline{\nabla}\equiv \nabla/h$, the entire velocity field is normalized by reference velocity $U_0$, i.e., $\mathcal{U}=(U,V)=(u,v)/U_0$, pressure is scaled as $\mathcal{P}=p/(\mu U_0/h)$, and the ionic concentrations are normalized as $\widetilde{C}_k=c_k/c_0$, where $c_0=(c_R+c_L)/2$ is the reference concentration. It is worth mentioning here that the dimensionless Debye parameter $\kappa_0h$, appearing in equations (\ref{num_eq}), is considered constant in the numerical simulations. However, the Reynolds number and P\'eclet number are taken as the same as obtained in analytical derivation. The above non-linear coupled equations (\ref{num_eq}) are modeled using the stabilized convection-diffusion equations and Laminar flow modules available in COMSOL platform. The transport equations (\ref{num_eq}) are solved using the suitable sets of boundary conditions as depicted in figure~\ref{fig_bc}(a). 

For the numerical simulations, we consider the quadratic shape function for electric potential, linear shape function for ion concentrations, second-order elements for the velocity, and linear elements for pressure. We check the grid independence of the model following the grid convergence index (GCI) method. We used a structured grid distribution with the help of a dummy variable $\Updelta$, which is proportional to the dimension of the model in both directions. Varying the value of $\Updelta$ from $2^0$ to $2^2$, we can achieve finer grid distributions consisting of doubled and quadrupled numbers of elements, respectively, for $\Updelta=2^1$ and $\Updelta=2^2$ as compared to $\Updelta=2^0$ case. Figure~\ref{fig_bc}(b) shows a typical grid distribution corresponding to $\Updelta=2^0$ for a model having smaller dimensions ($l=20h$, $R_h=R_l=15h$) as compared to the model used for the present analysis. In figure~\ref{fig_bc}(c), the grid convergence indexes $(GCI)$ are shown for two sets of grid refinements, i.e., $GCI_{12}=2.4\%$ and $GCI_{24}=0.3\%$, which also satisfy that results (here, volume flux) approach towards an asymptotic solution, i.e., $GCI_{12}/2^{p}GCI_{24}\approx1$ with an order of convergence $p=2.49$. Thus, following this exercise, we consider $\Updelta=2^2$ in the present study for further calculations demonstrated in the forthcoming sections. Moreover, the accuracy of the developed numerical model is further ascertained by verifying that the model is free from any spurious flow in the absence of the external electric field and zero concentration gradient. Also, an attempt has been made to ensure that the results of this analysis are not biased toward the reservoir size. For this part, we performed the `reservoir size effect' analysis and established that $(50h\times50h)$ is the optimum size for this analysis (cf. figure~\ref{fig_bc}d). In the present study, until mentioned otherwise, all the results are derived for height $h=20$nm, length $l=100\times h=2\mu$m, reservoir size $R_h=R_l=50\times h=1\mu$m.

\section{Discussion of the results}

\subsection{Selection of model parameters}

For the results in this article, we initially consider the very small magnitude of surface charge density ($\leq 2$ mC/m$^2$). This is considered to employ the Debye-H\"uckel approximation for the entire range of bulk concentrations considered in obtaining the results both analytically and numerically, as presented in the forthcoming sections. The bulk concentrations in the reservoirs are varied within a range of $\sim1$ mM to $\sim150$ mM, which covers the typical range of bulk concentrations usually considered in the practical nanofluidic application \citep{kim2007concentration,he2017electrokinetic}. In figure~\ref{fig_result_1}, we have considered the concentration difference $|\Updelta c|\leq 50$ mM, whereas this difference is created with the fixed concentration of $c_L=100$ mM at the left reservoir. The reason for having such a maximum bound of $\Updelta c$ is essentially to maintain $\Updelta c/c_0<1$, a condition necessary to obtain the solutions in the analytical framework. In the next result (figure~\ref{fig_result_2}), where the influence of surface charge density is investigated at a constant $\Updelta c$, we consider $\Updelta c=-50$ mM because it provides the maximum chemiosmotic flow (COF) velocity as witnessed in figure~\ref{fig_result_1}.

\subsection{Non-overlapping EDL}

\begin{figure}
  \centering
  \subfigure{\includegraphics[width=1.7in]{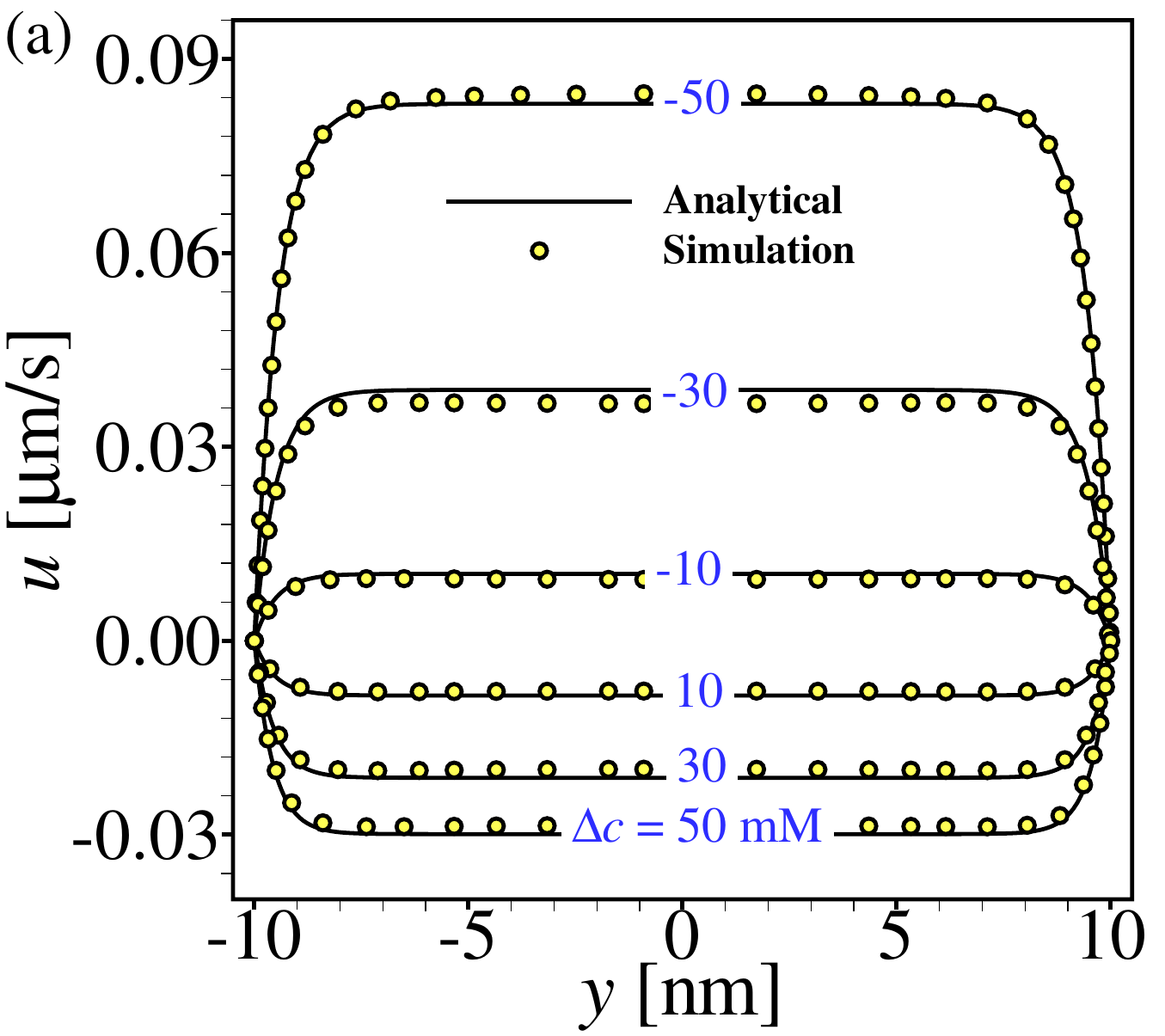}}
  \subfigure{\includegraphics[width=1.7in]{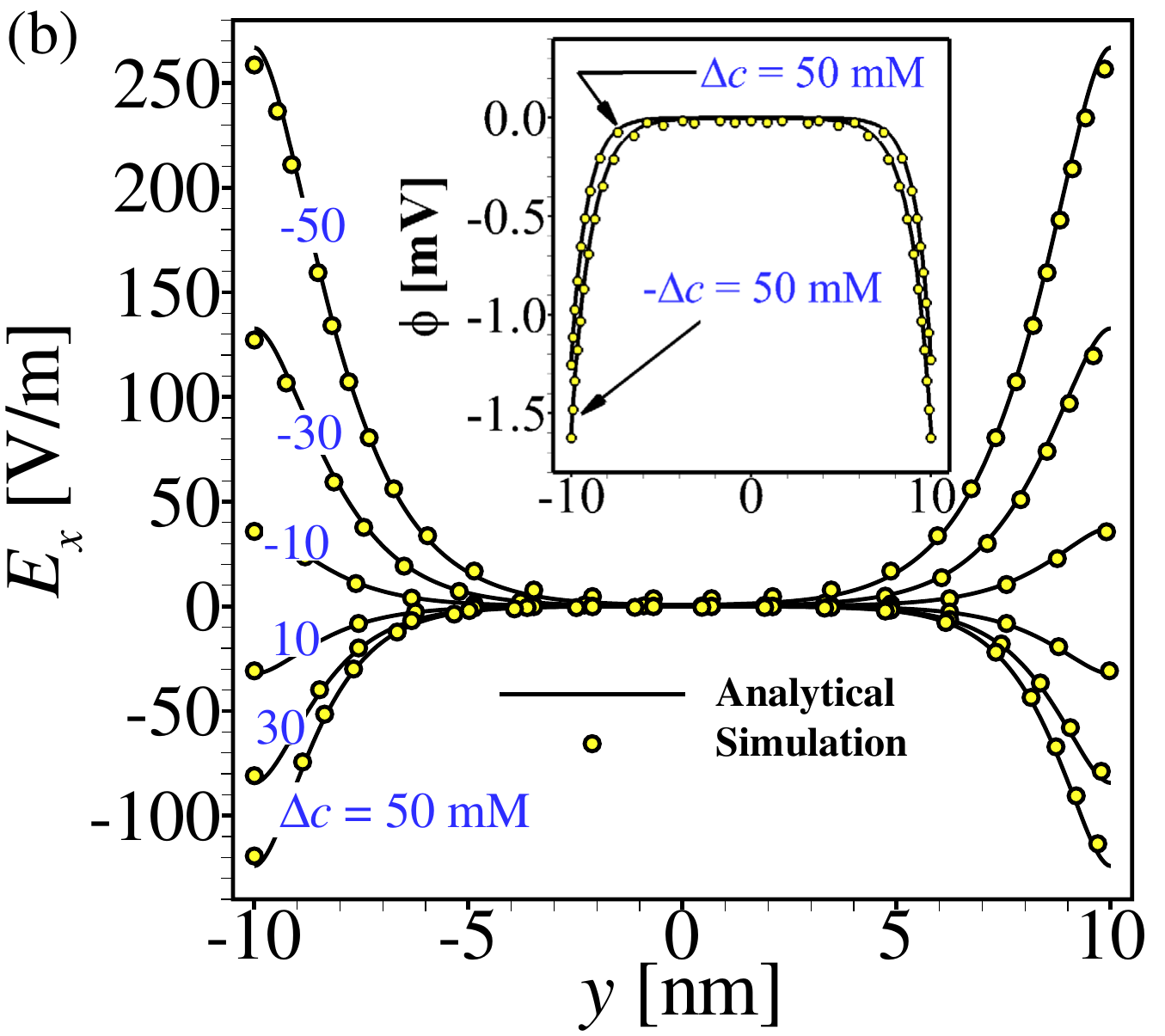}} 
  \subfigure{\includegraphics[width=1.7in]{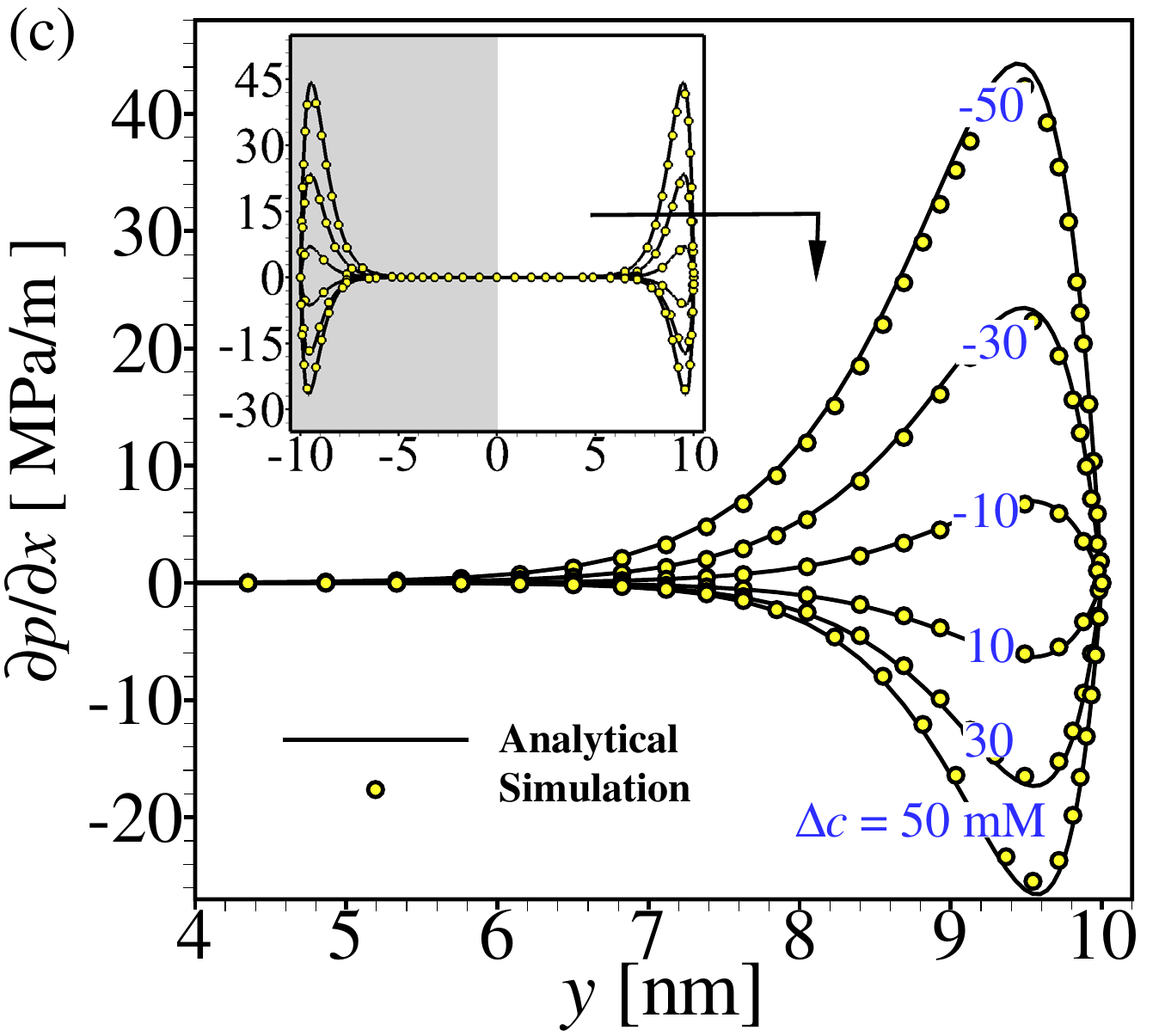}}
  \subfigure{\includegraphics[width=1.7in,height=1.5in]{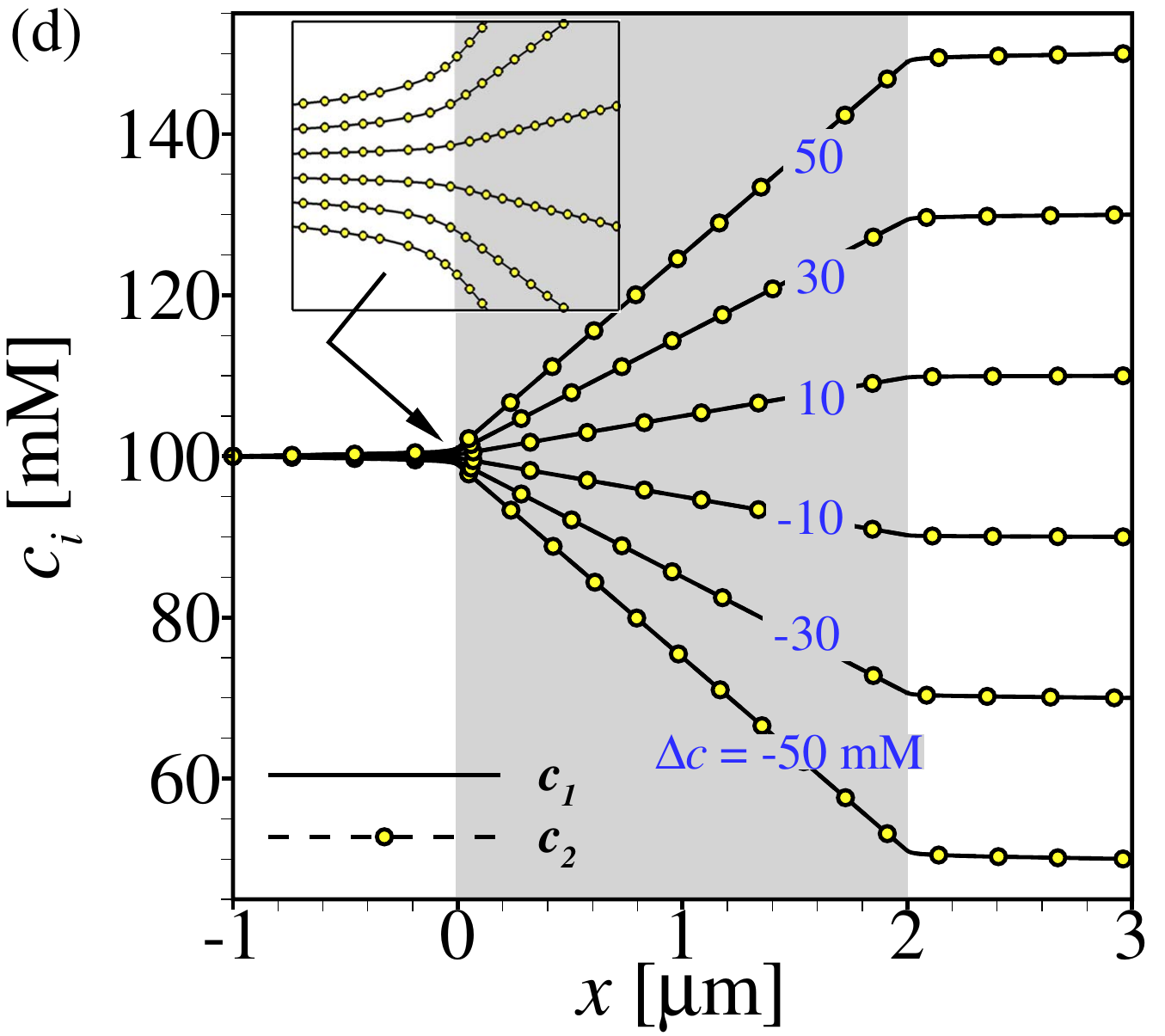}}  
  \subfigure{\includegraphics[width=3.4in,height=1.6in]{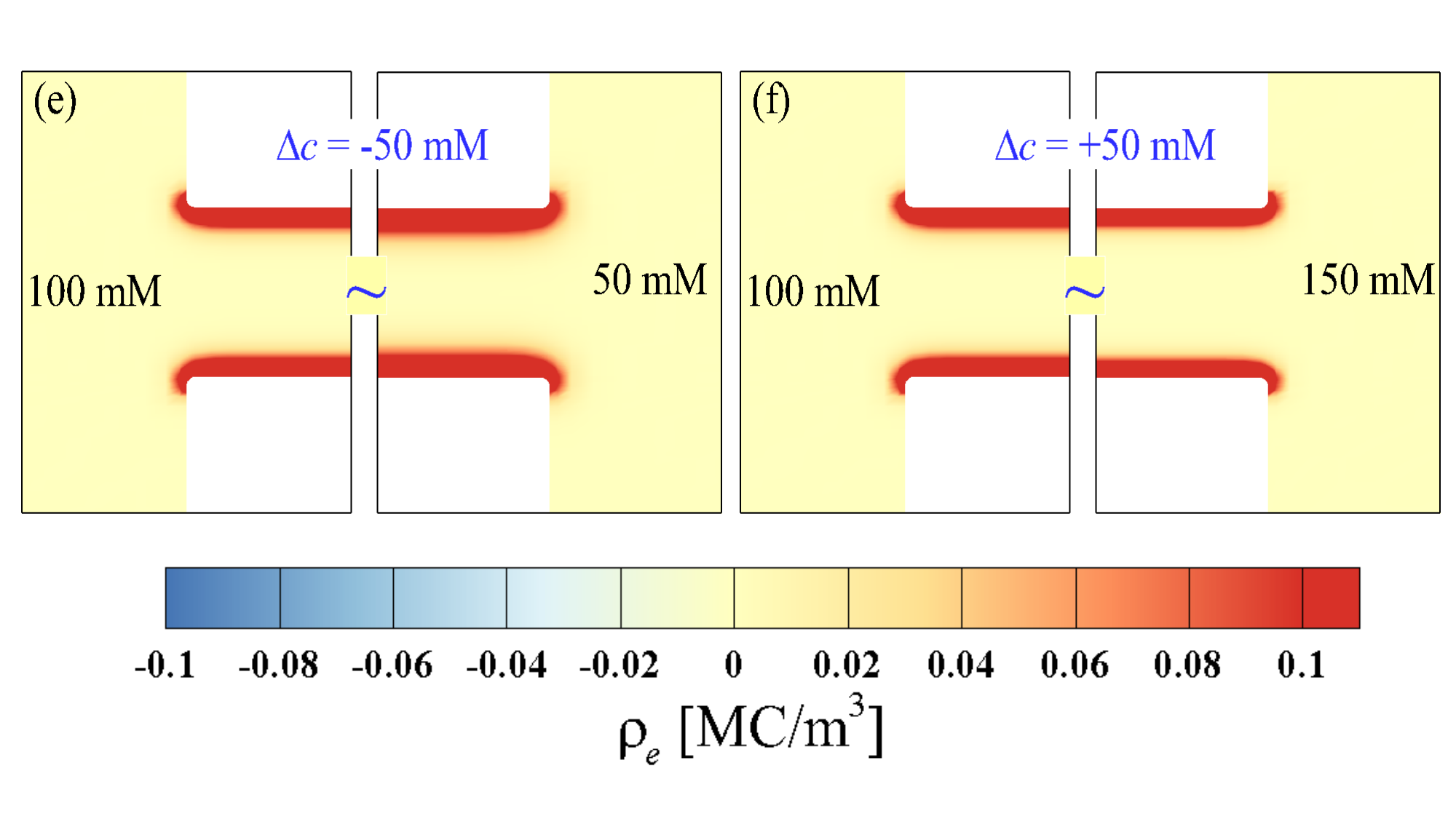}}
  \caption{(a) Axial COF velocity, (b) axial electric field with electric potential (inset), (c) pressure gradients across nanochannel cross-section and (d) concentrations of ions along $y=0$ for different values of $\Updelta c=\pm10,\pm30,\pm50$ mM at $\sigma_t=\sigma_b=-1$ mC/m$^2$, $c_L=100$ mM, $c_R=c_L+\Updelta c$. Solid lines represent the analytical solution for axial velocity Eq.(\ref{NS_eq_9}), electric potential Eq.(\ref{P_equ_5}) and axial electric field Eq.(\ref{P_equ_6}), and the corresponding numerical solutions are denoted by symbols. (e,f) Represents the corresponding space charge density in the vicinity of the nanochannel openings for $\Updelta c=\pm50$ mM.} 
  \label{fig_result_1}
\end{figure}

We start our discussions with the axial COF velocity profile as well as corresponding electric potential distribution, axial electric field, axial pressure gradient, and bulk ionic concentrations obtained for different values of $\Updelta c=\pm10,\pm30,\pm50$ mM at $\sigma_t=\sigma_b=-1$ mC/m$^2$, as depicted in figure~\ref{fig_result_1}a-d. Since no external potential is applied, the axial electric field is only generated due to the non-vanishing axial gradient present in the EDL potential. Thus, the axial electric field $E_x$ vanishes outside the EDL. It is presented in figure~\ref{fig_result_1}e,f that the thickness of the EDL gradually changes along the axial direction. This observation is attributed to the effect that stems from the axially varying bulk concentration. Figure~\ref{fig_result_1}e,f shows that the thickness of the EDL gradually increases towards the reservoir having a comparatively lower bulk concentration. As demonstrated in figure~\ref{fig_result_1}b, we can observe that the magnitude of the axial electric field is proportional to the magnitude of the concentration difference across the reservoir. On the contrary, the direction of this axial electric field directly depends upon the sign of local wall surface charge density and the direction of the concentration gradient. This observation is witnessed in figure~\ref{fig_result_1}b as well. Similarly, in the absence of any external pressure difference, the local axial pressure gradient is also generated due to the concentration difference. Thus, it takes the maximum value within the EDLs and starts to disappear in bulk, i.e., at $Y=0$ (figure~\ref{fig_result_1}c).

We observe from figure~\ref{fig_result_1}a that the direction of axial COF velocity within the nanochannel depends upon the direction of the concentration gradient. More specifically, the direction of the axial COF velocity within the nanochannel is in the opposite direction of the concentration gradient. This observation underlines that the COF, like the normal diffusion process, is directed from a higher concentration toward a lower concentration. It is quite intuitive and also witnessed in figure~\ref{fig_result_1}a that the magnitude of COF axial velocity within the nanochannel is proportional to the corresponding concentration difference created between both reservoirs. However, as depicted in figure~\ref{fig_result_1}a, the magnitude of axial COF velocity for a given value of $\Updelta c$ is not the same if the direction of $\Updelta c$ changes, i.e., the magnitude of axial velocity is not the same for any given $\pm\Updelta c$. Figure~\ref{fig_result_1}a demonstrates that for the identical magnitude of $|\Updelta c|$, the axial COF velocity for $\Updelta c<0$ is higher than that for $\Updelta c>0$. Here, the bulk (reference) concentration $c_0$ reduces for $\Updelta c<0$ as compared to $\Updelta c>0$ since the concentration at the left reservoir is always kept fixed. This indicates that the magnitude of axial COF velocity is inversely proportional to the bulk concentration, which is similar in nature as compared to the normal EOF under identical configuration. The underlying reason behind this observation is as follows. With increasing the magnitude of bulk concentration, the surface potential reduces, which in turn, reduces the magnitude of COF velocity.

\begin{figure}
  \centering
  \subfigure{\includegraphics[width=1.7in]{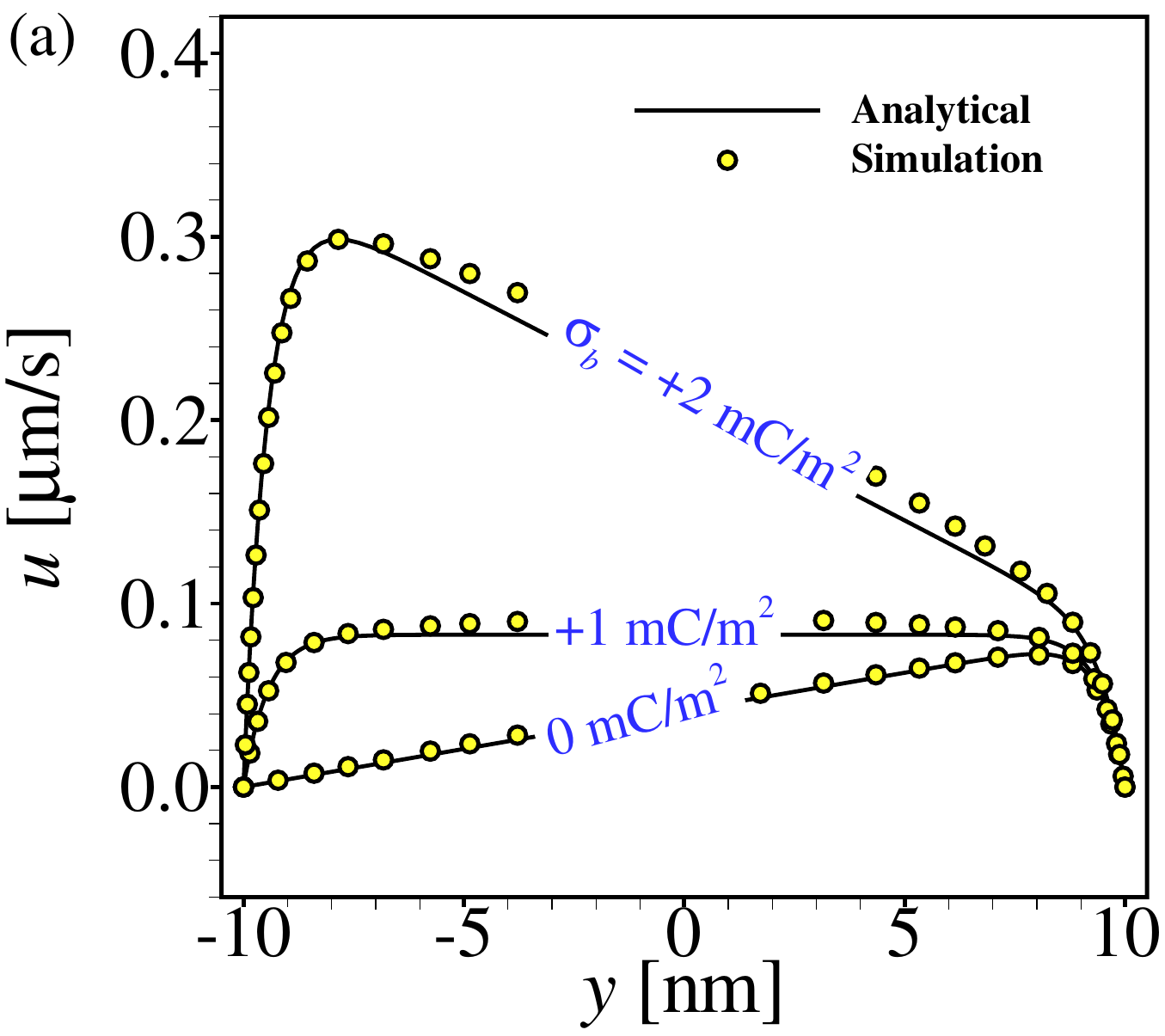}}
  \subfigure{\includegraphics[width=1.7in]{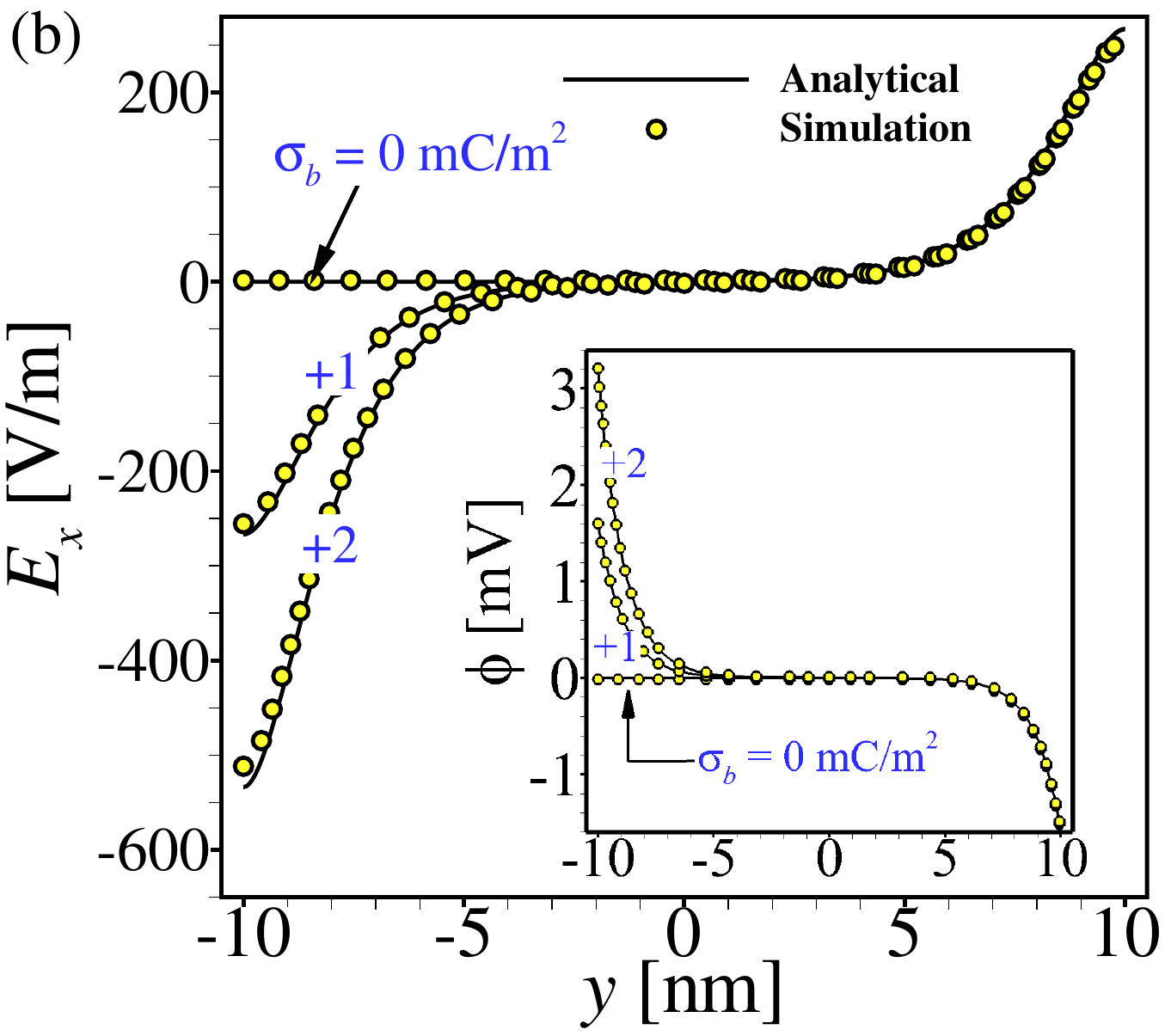}} 
  \subfigure{\includegraphics[width=1.7in]{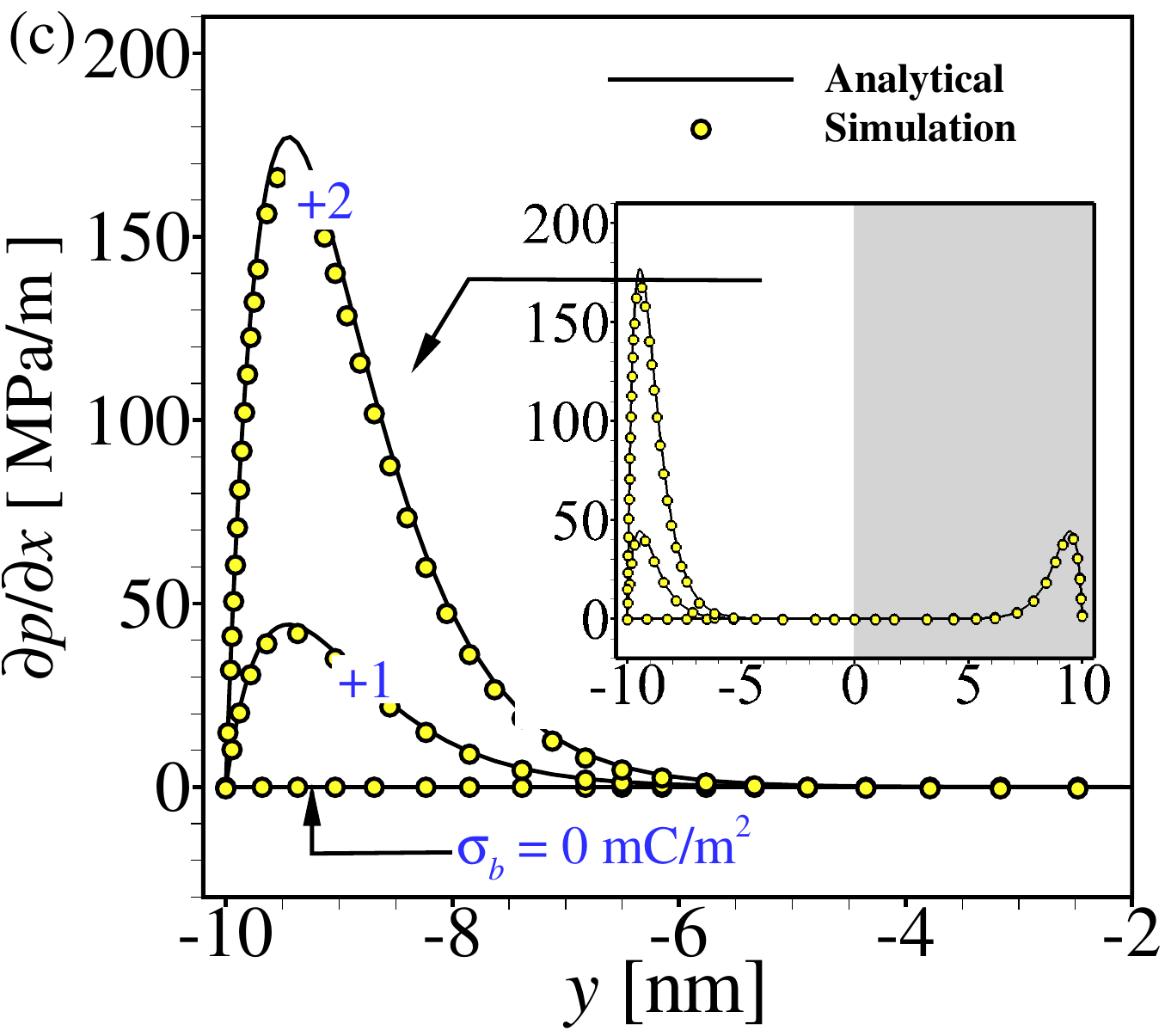}}
  \subfigure{\includegraphics[width=1.9in]{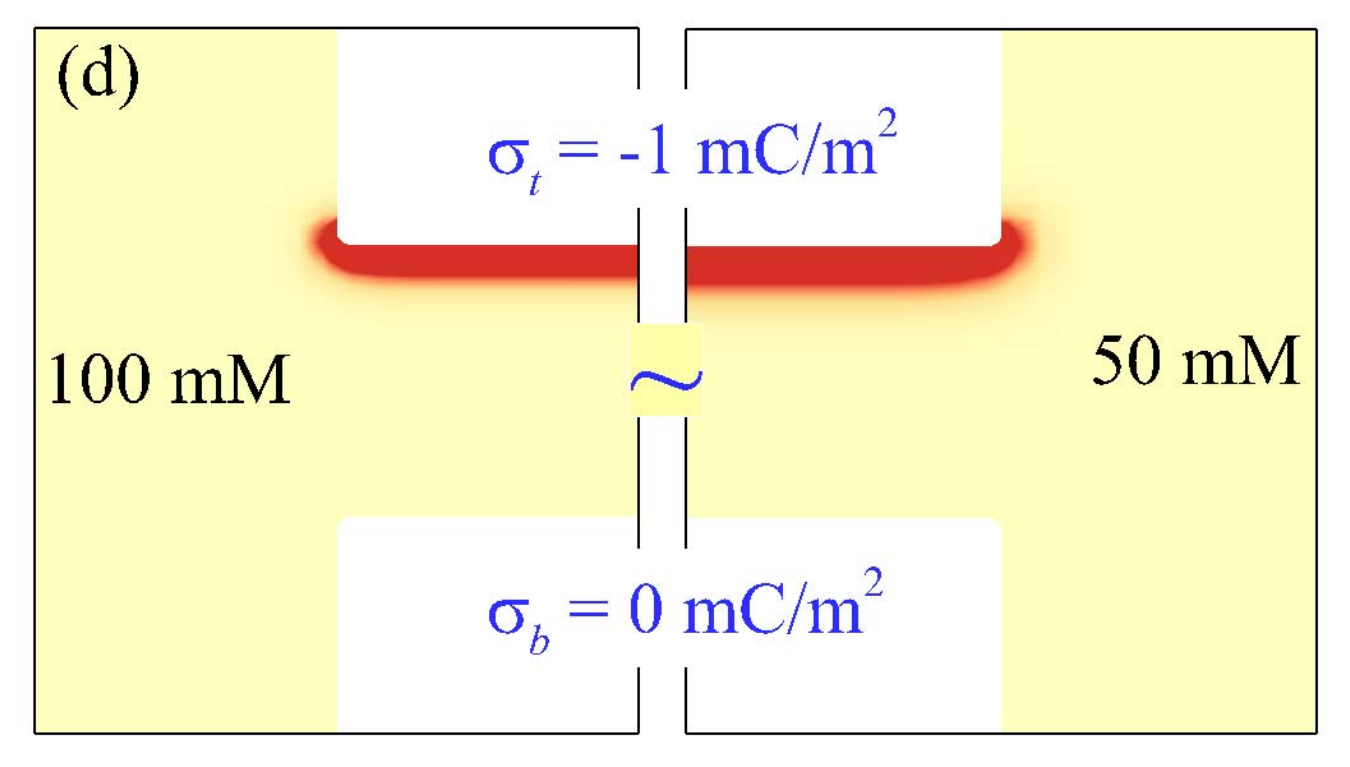}}  
  \subfigure{\includegraphics[width=1.9in]{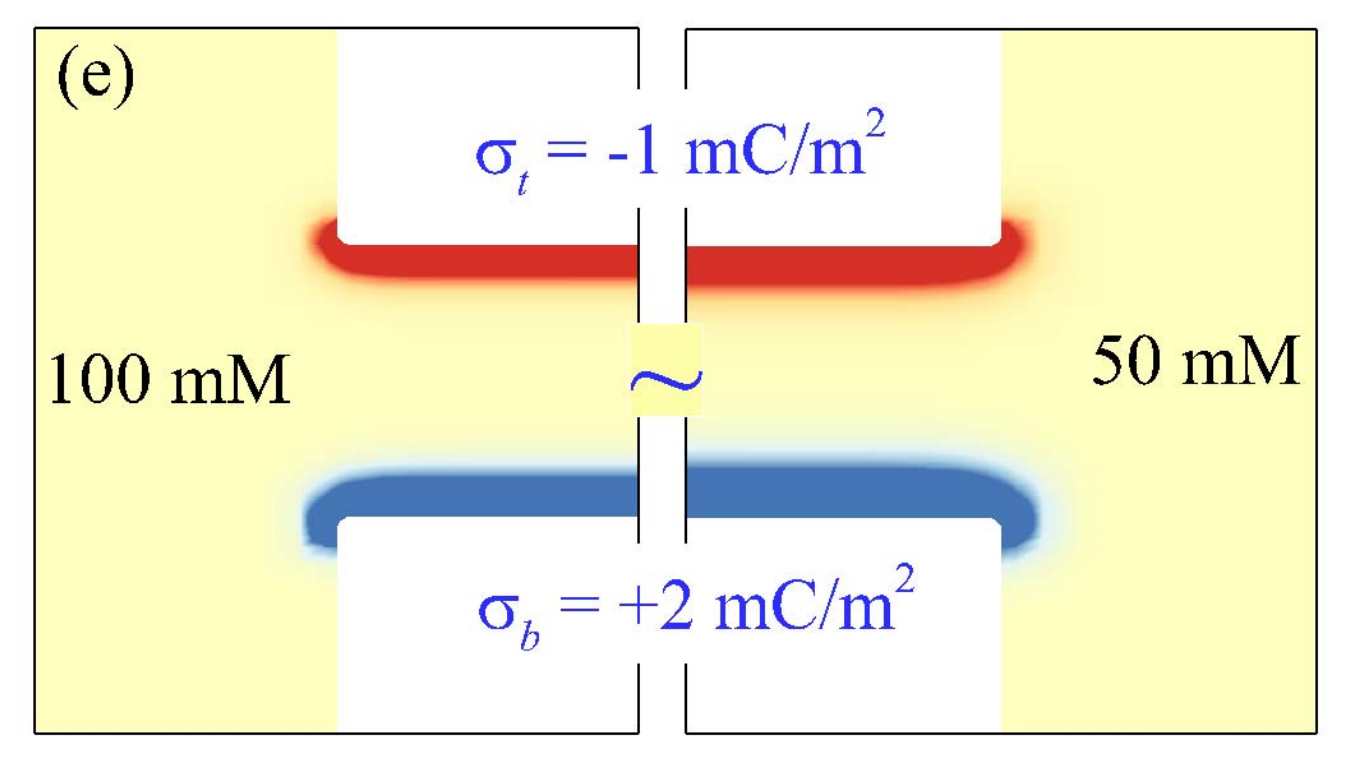}}
  \subfigure{\includegraphics[width=0.6in]{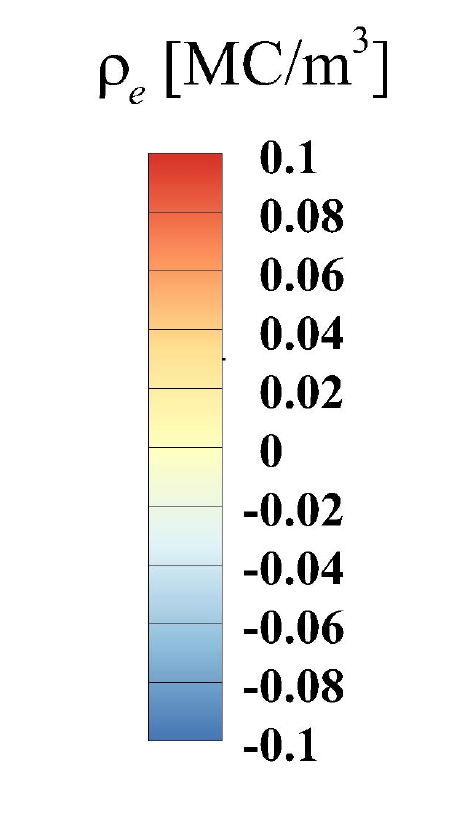}}
  \caption{(a) Axial COF velocity, (b) axial electric field with electric potential (inset), (c) pressure gradients across nanochannel cross-section for different values of $\sigma_b=0,+1,+2$ mC/m$^2$ at $\Updelta c=-50$ mM, $\sigma_t=-1$ mC/m$^2$, $c_L=100$ mM, $c_R=50$ mM. Solid lines represent the analytical solution for axial velocity Eq.(\ref{NS_eq_9}), electric potential Eq.(\ref{P_equ_5}) and axial electric field Eq.(\ref{P_equ_6}), and the corresponding numerical solutions are denoted by symbols. (d,e) Represents the corresponding space charge density in the vicinity of the nanochannel openings for $\sigma_b=0,+2$ mC/m$^2$.} 
  \label{fig_result_2}
\end{figure}

Figure~\ref{fig_result_2}a shows the impact of surface charge density on the axial COF velocity at $\sigma_t=-1$ mC/m$^2$, $\sigma_b=0,+1,+2$ mC/m$^2$, and $\Updelta c=-50$ mM. Unlike EOF, we find that the direction of axial COF velocity does not depend on the sign of wall charge density. As witnessed in figure~\ref{fig_result_2}a, the direction of axial COF velocity remains unaltered even if the sign of $\sigma_b$ and $\sigma_t$ is opposite. Moreover, at $\Updelta c=-50$ mM, the axial velocity obtained for $\sigma_t=-\sigma_b=-1$ mC/m$^2$ and $\sigma_t=\sigma_b=-1$ mC/m$^2$ is exactly the same (cf. figure~\ref{fig_result_1}a and figure~\ref{fig_result_2}a). \textit{To explain this interesting phenomenon, we plotted the corresponding axial electric field and axial pressure gradient in figures~\ref{fig_result_2}b and c, respectively.} Since no external electric field is applied, the axial electric field develops only due to the axial variation of the local space charge density. As witnessed in figure~\ref{fig_result_2}d,e, the EDL gradually becomes thicker along the $x-$axis, and this observation is attributed to the effect of $\Updelta c<0$ on the underlying interfacial electrochemical phenomenon. Thus, at $\Updelta c<0$, the direction of the axial electric field $E_x(-\partial_X\Psi)$ is exactly opposite of the sign of the local wall charge density. Also, the sign of local space charge density $\rho_e$ always has the opposite sign of the corresponding wall charge density. So, the direction of axial COF velocity ($E_x\times\rho_e>0$) under a negative concentration difference $\Updelta c<0$ remains positive irrespective of any sign of surface charge density. From the forging discussion, we can conclude that, unlike EOF, the direction COF is invariant to the sign of local surface charge density.

Figure~\ref{fig_result_2}a also demonstrates that the magnitude of axial COF velocity is proportional to the magnitude of local wall charge density. Thus, it vanishes in the absence of local wall charge density at $\sigma_b=0$ mC/m$^2$ (cf. figure~\ref{fig_result_2}d). As witnessed in figure~\ref{fig_result_1}a and figure~\ref{fig_result_2}a, despite different signs of surface charge density $(\pm\sigma_b=\sigma_t=-1$ mC/m$^2$), the axial COF velocity takes the infamous plug-like shape if the magnitude of the surface charge densities is the same, and the bulk concentration is sufficiently higher. Moreover, for a fixed concentration gradient $\Updelta c=-50$ mM, as considered here, the velocity obtained with $\sigma_b=-\sigma_t$ is exactly the same as derived for $\sigma_b=\sigma_t$ (cf. figures~\ref{fig_result_1}a and \ref{fig_result_2}a). Furthermore, the axial COF velocity becomes asymmetric with respect to the nanochannel axis for the unequal magnitude of surface charge densities specified at both walls. Quite notably, as seen in figure~\ref{fig_result_2}a, the analytical solutions match well with the numerical results for all cases under Debye-H\"uckel approximation.

\begin{figure}
  \centering
  \subfigure{\includegraphics[width=2.2in]{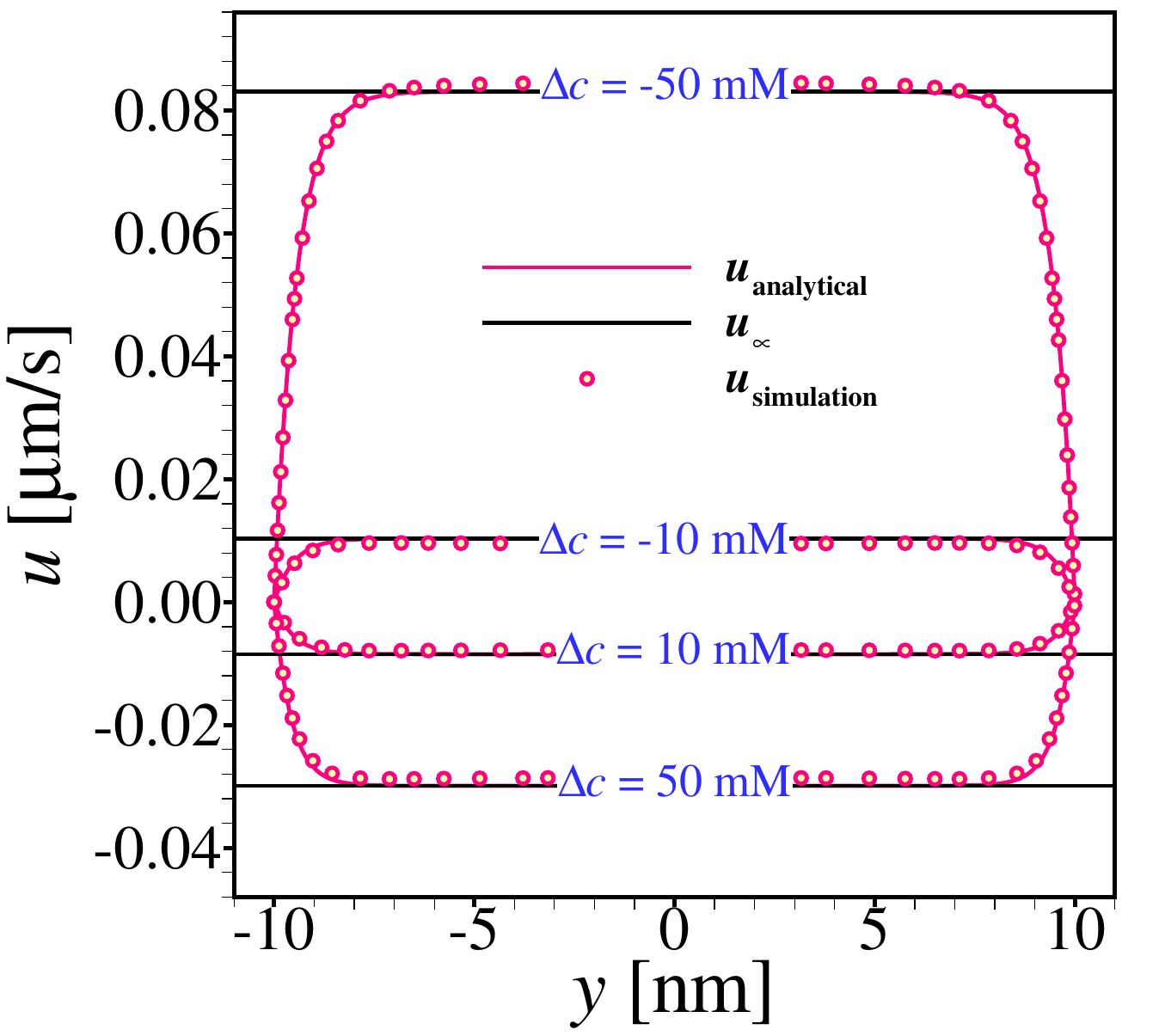}}
  \caption{Axial COF velocity across nanochannel cross-section for different values of $\Updelta c=\pm10,\pm30,\pm50$ mM at $\sigma_t=\sigma_b=-1$ mC/m$^2$, $c_L=100$ mM, $c_R=c_L+\Updelta c$. Solid black lines represent the analytical solution for axial velocity Eq.(\ref{NS_eq_9}), and symbols denote the corresponding numerical solutions. The solid black lines denote the effective COF slip velocity $u_\infty^{COF}$ as presented in eq.(\ref{slip velocity}).} 
  \label{fig_result_3}
\end{figure}

We observe that the axial COF velocity takes the plug-like shape like EOF under a non-overlapping EDL scenario for the identical magnitude of the surface charge densities at both walls. This observation opens up the scope for obtaining the effective axial COF velocity scale $(u_\infty^{COF})$ over a charged surface under non-overlapping EDL. In order to obtain the effective COF velocity scale $u_\infty^{COF}$, we integrate Eq.(\ref{NS_eq_7}) in the absence of external pressure gradient $(d_XP_0=0)$, and external electric field $(\Bar{E}=0)$ with respect to the $Y-$ coordinate from the charged surface to infinity. Under the framework of Debye-H\"uckel approximation, considering no-slip boundary condition at the charged wall and vanishing lateral gradient $(\partial_YU=0)$ at infinity, we obtain the dimensional effective axial COF velocity scale as
\begin{equation}
u_{\infty}^{COF}=-\frac{\nabla c\sigma^2\phi_0}{16\mu F c_0^2}.
\label{slip velocity}
\end{equation}
Here, $\sigma$ denotes the dimensional surface charge density, $\phi_0(=k_BT/e)$ refers to the thermal potential, $\mu$ represents the viscosity of the electrolyte, and $F$ denotes the Faraday constant. It may be mentioned here that $c_0=(c_L+c_R)/2$ represents the reference concentration, which is also the average concentration of the two different concentrations provided at both ends. The expression of effective COF velocity appears in Eq.(\ref{slip velocity}) shows that the direction of COF is opposite to the direction of concentration gradient $(\nabla c)$ and independent of the sign of surface charge density as realized by $\sigma^2$. It is also evident from Eq.(\ref{slip velocity}) that the magnitude of the effective COF is proportional to the concentration gradient. It is worth mentioning here that this aspect is similar to that of the electroosmotic flow (EOF), which is proportional to the driving strength axial electric field. However, unlike EOF, the COF can be enhanced square-folded by increasing the surface charge density. Also, the axial COF velocity is inversely proportional to the square of bulk concentration. This justifies the reason behind not obtaining the same magnitude of COF velocity for $\pm\Updelta c$ as presented in figure~\ref{fig_result_1}a.  Figure~\ref{fig_result_3} shows that the effective COF slip velocity $u_\infty^{COF}$ matches pretty well with both analytical and numerical solutions of COF velocity outside the EDL region. 

For sufficiently lower values of specified zeta potential ($\zeta=\sigma\lambda_D/\epsilon_0\epsilon_r$) at the channel wall, the COF slip velocity from Eq.(\ref{slip velocity}) can be represented as
\begin{equation}
 u_{\infty}^{COF}=-\frac{\epsilon_0\epsilon_r\nabla c \zeta^2}{8\mu c_0} .
\end{equation}
Comparing the above COF slip velocity with the well-established EOF slip velocity $(u_\infty^{EOF}=-\epsilon_0\epsilon_r E\zeta/\mu)$ one can obtain the following
\begin{equation}
    \frac{u_\infty^{EOF}}{u_\infty^{COF}}=8\frac{E/\zeta}{\nabla c/c_0}.
    \label{E2C}
\end{equation}
Under the framework of the Debye-H\'uckel limit and weak field assumption, both the external electric field in EOF and the concentration gradient in COF are considered to be sufficiently small to withstand the local equilibrium of the ion cloud in the vicinity of the charged wall, i.e., $E/\zeta<1 $ and $\nabla c/c_0<1$. It is worth mentioning here that the above Eq.(\ref{E2C}) implies a relation between the electroosmotic and chemiosmotic slip velocities for low surface potential, i.e., $u_\infty^{EOF}=8\times u_\infty^{COF}$ if $E/\zeta\approx\nabla c/c_0$. This scaling analysis is suggestive that pertaining to the cases of low zeta potential, the EOF slip velocity is approximately one order higher than the corresponding COF slip velocity.

\subsection{Overlapping EDL}
\begin{figure}
  \centering
  \subfigure{\includegraphics[width=1.25in]{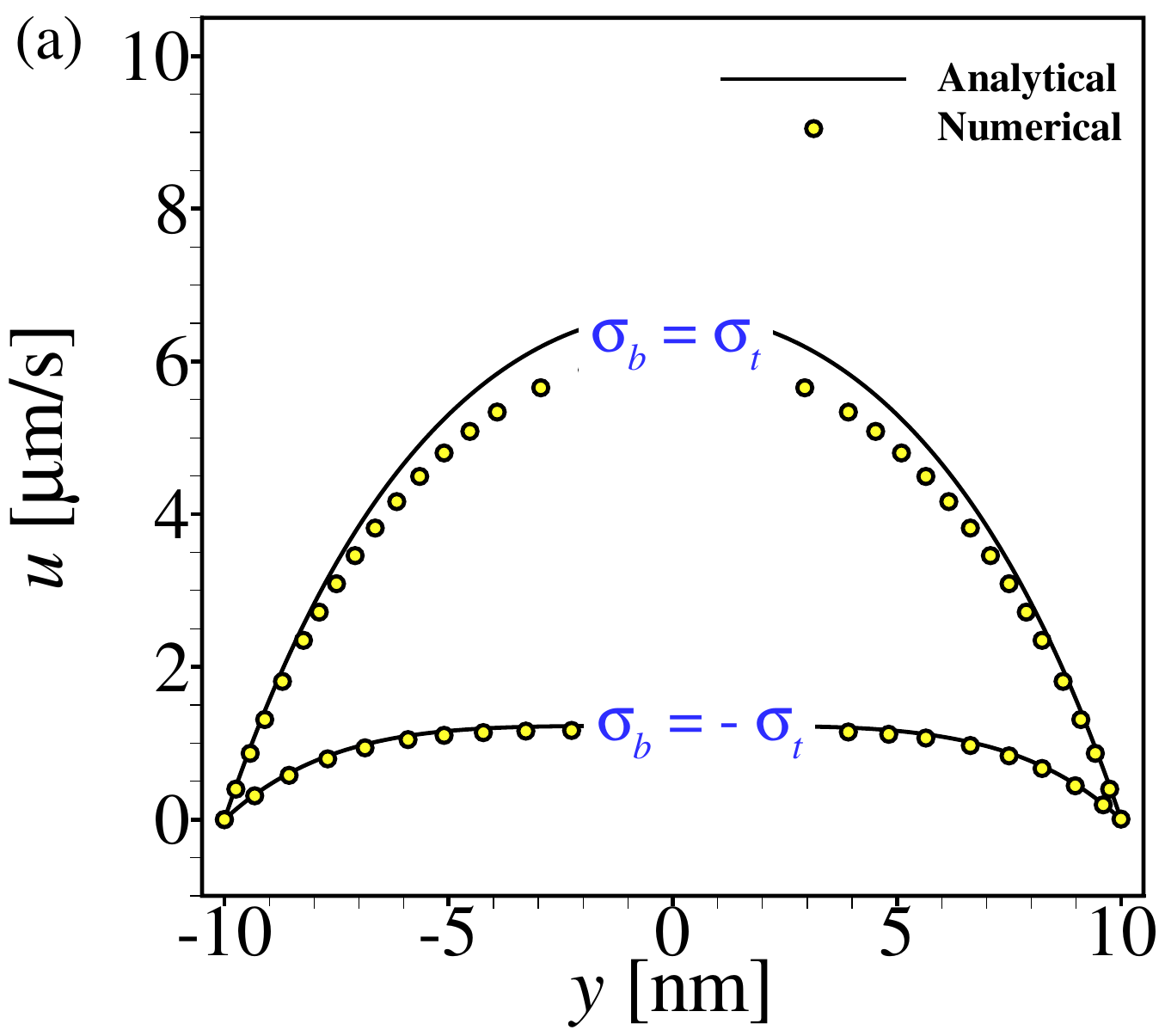}}
  \subfigure{\includegraphics[width=1.25in]{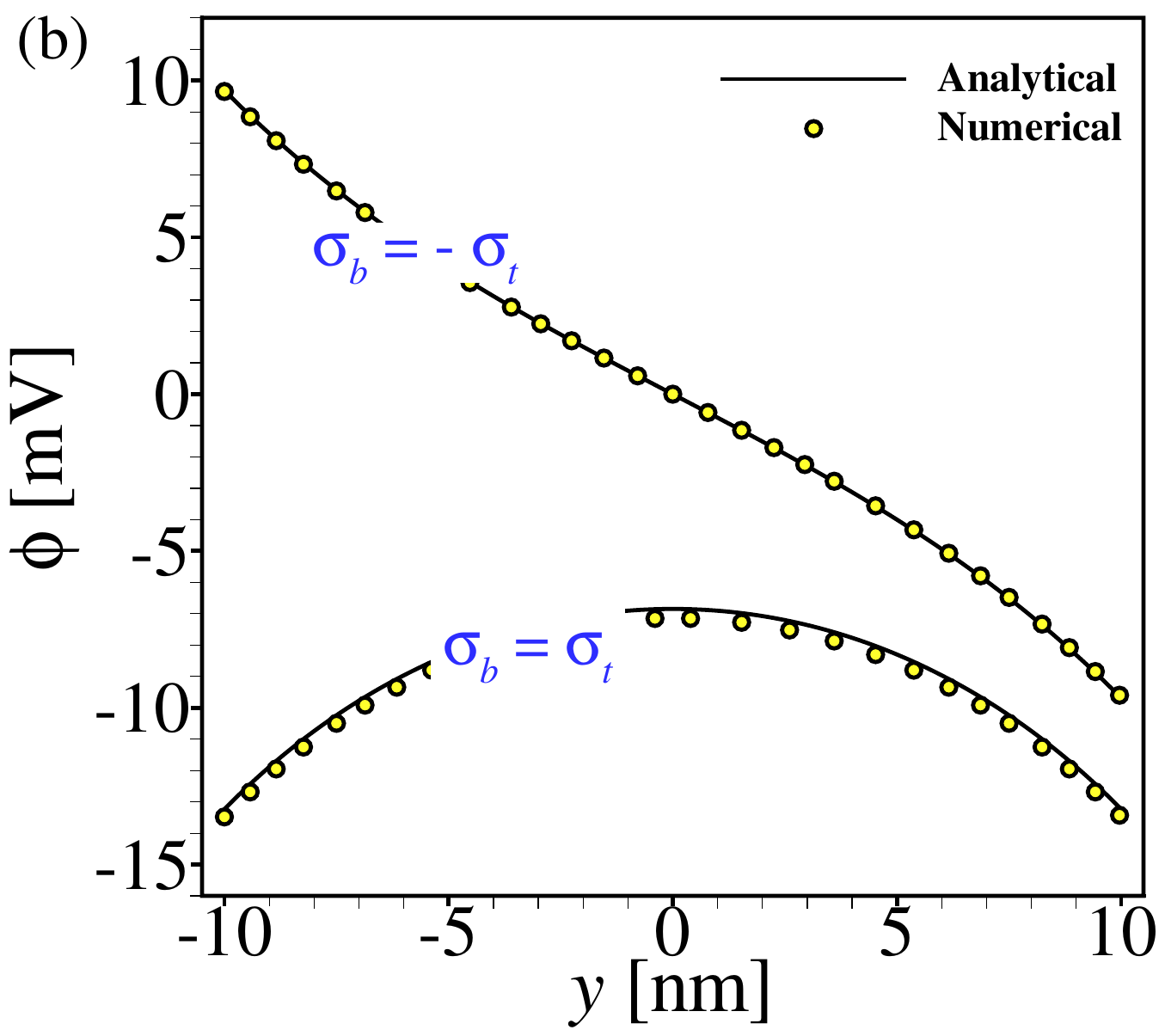}} 
  \subfigure{\includegraphics[width=1.25in]{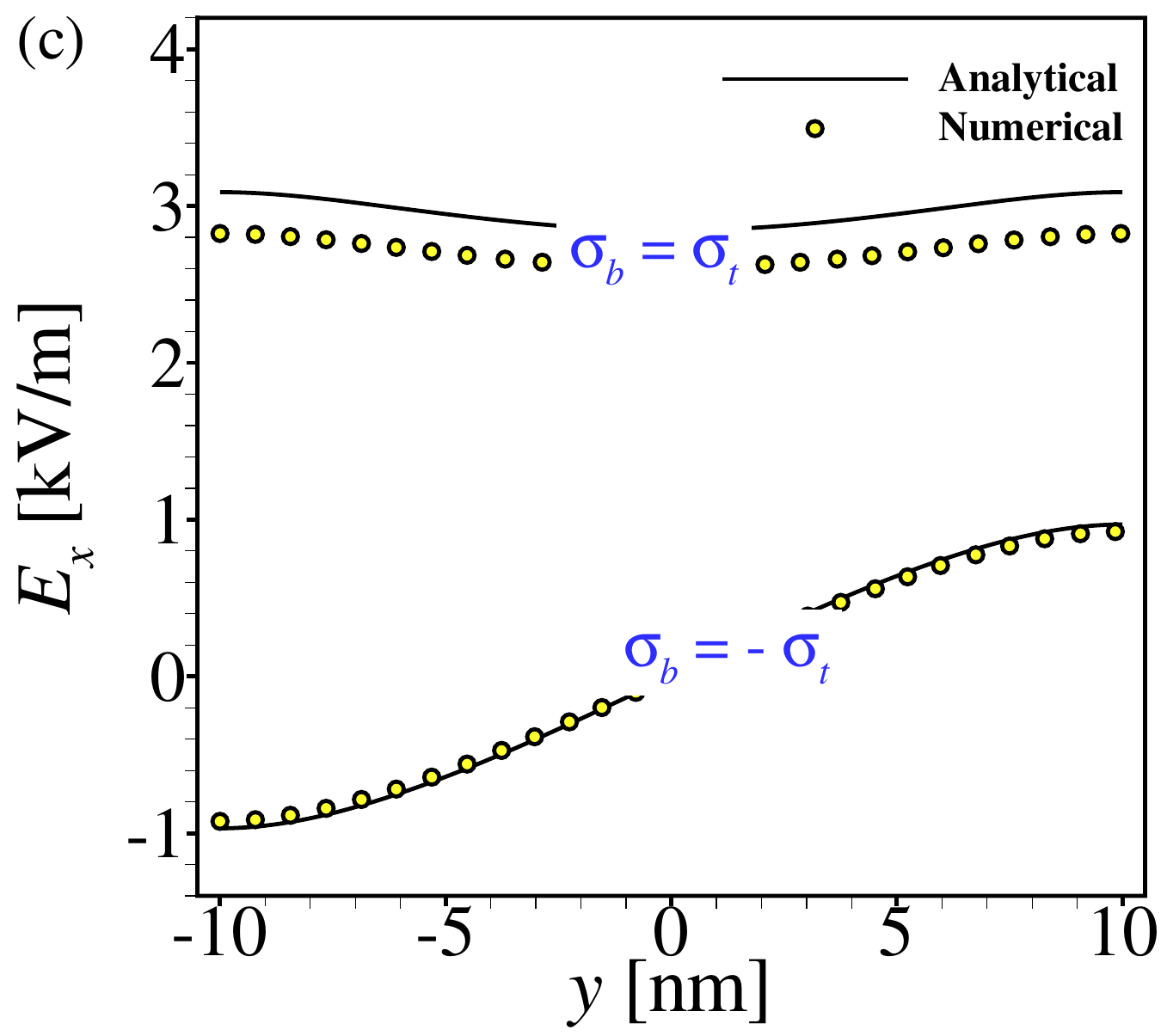}}
  \subfigure{\includegraphics[width=1.25in]{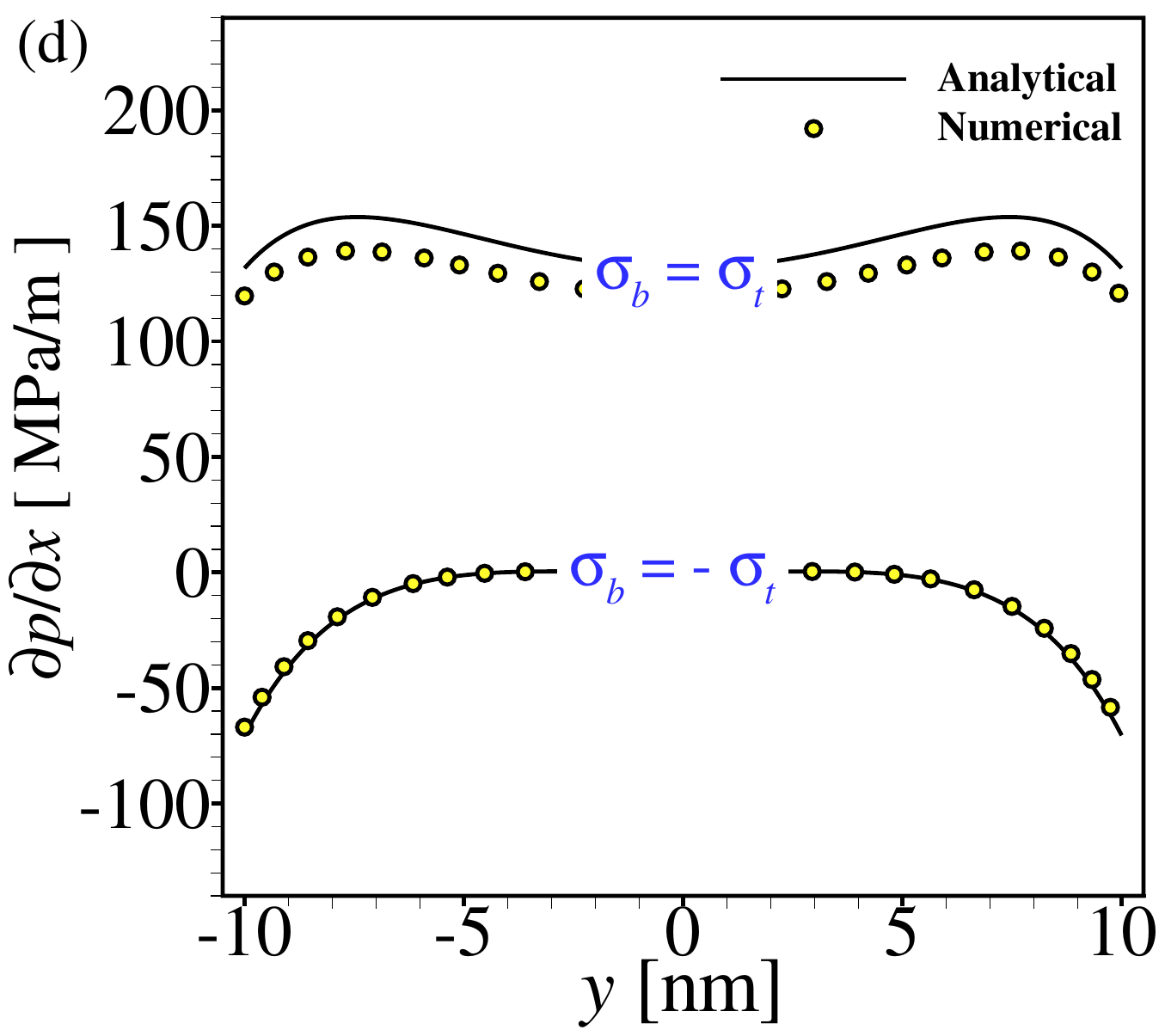}} \\
  \subfigure{\includegraphics[width=1.9in]{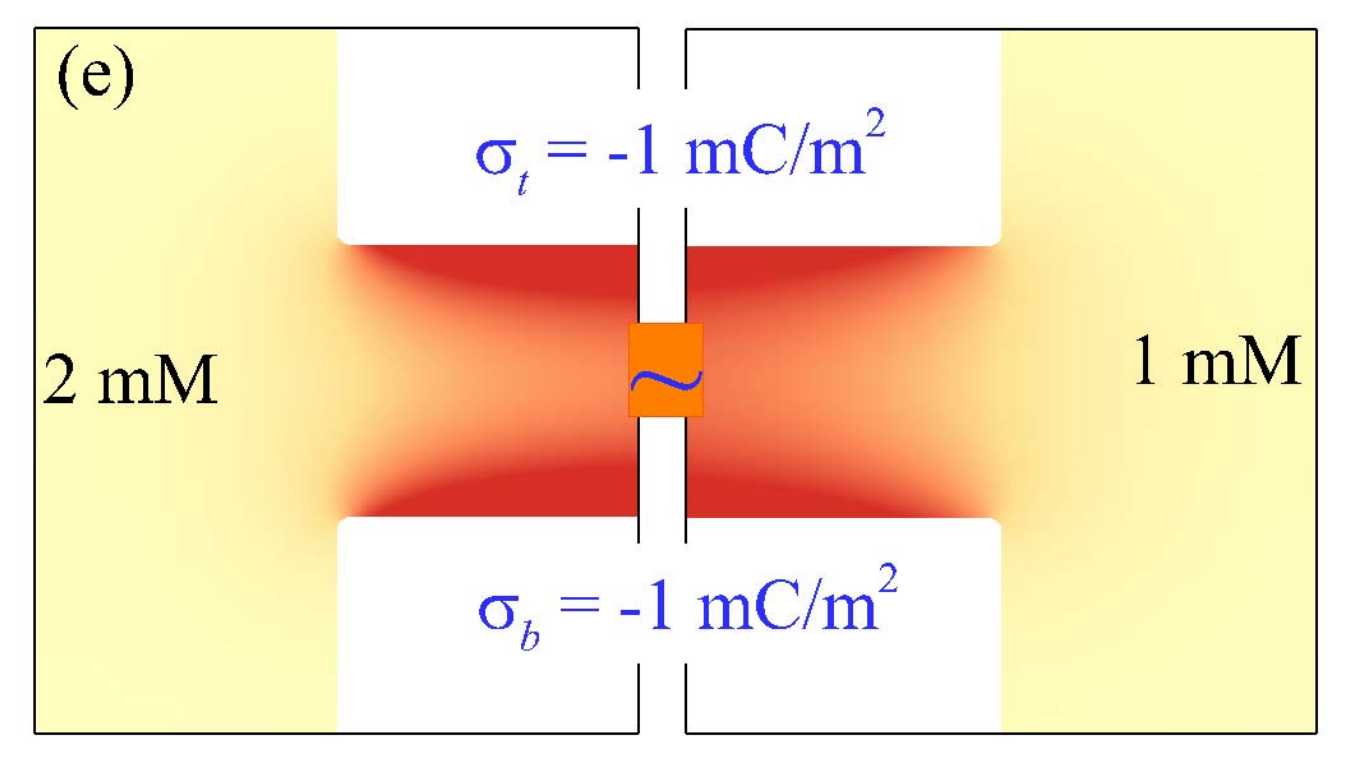}}
  \subfigure{\includegraphics[width=1.9in]{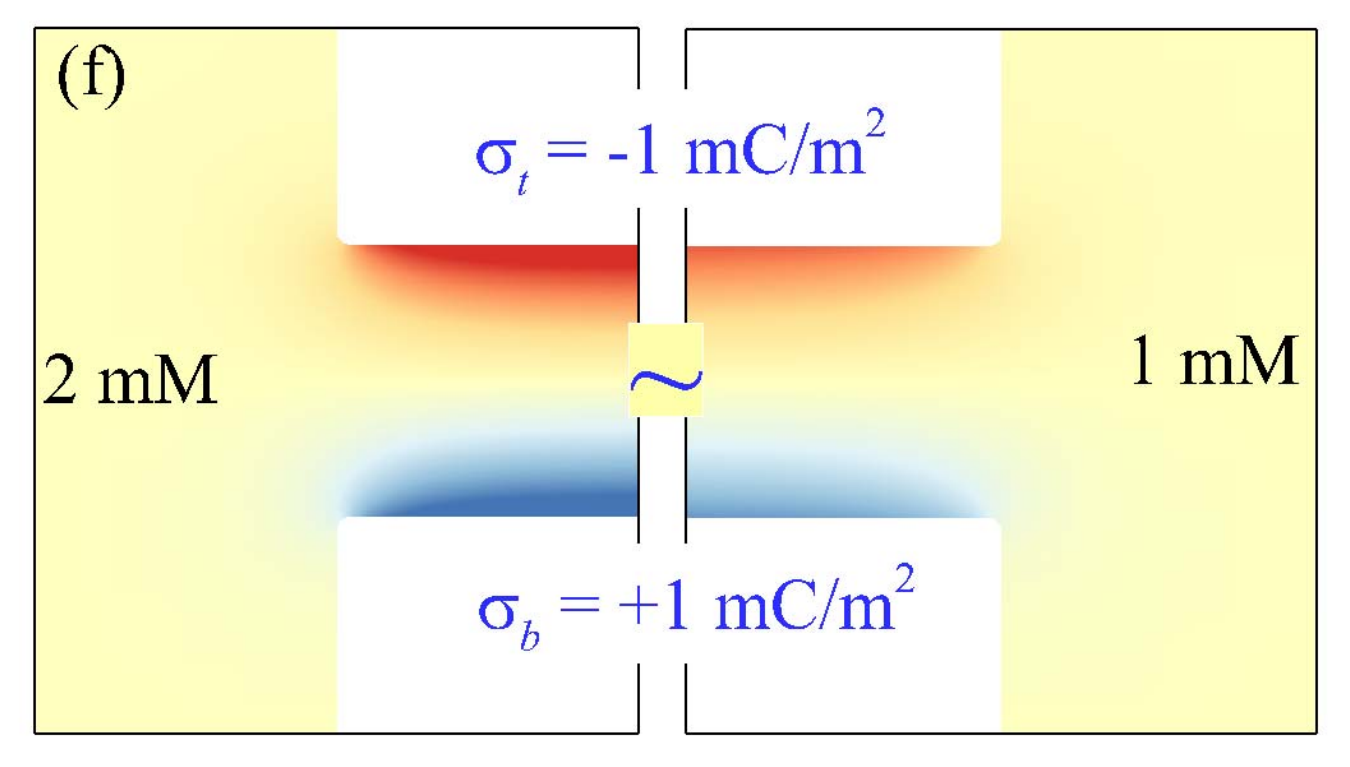}}
  \subfigure{\includegraphics[width=0.6in]{r2_f.pdf}}
  \caption{(a) Axial COF velocity, (b) electric potential, (c) axial electric field, and (d) axial pressure gradient across the nanochannel cross-section; (e,f) space charge density in the vicinity of channel openings for $\sigma_b=\sigma_t=-1$ mC/m$^2$ and $-\sigma_b=\sigma_t=-1$ mC/m$^2$ at $c_L=2$ mM, $c_R=1$ mM.} 
  \label{fig_result_4}
\end{figure}

Results presented in the previous subsection show that the axial COF velocities obtained for $\sigma_t=\sigma_b$ and $\sigma_t=-\sigma_b$ are identical for non-overlapping EDL cases. However, the underlying transport feature of COF may not follow the same behavior for overlapping EDL scenarios. To investigate that aspect in greater detail, we present the axial COF velocity and corresponding electric potential, axial electric field, and pressure gradient, respectively, in figures~\ref{fig_result_4}a,b,c, and d, obtained at a fixed $\Updelta c$ and fixed magnitude of surface charge density, i.e., $|\sigma_b|=|\sigma_t|=1$ mC/m$^2$. 

We observe in figure~\ref{fig_result_4}a that the velocity obtained for these two cases are not the same; rather the axial COF velocity obtained at $\sigma_t=\sigma_b$ is higher than the axial COF velocity obtained at $\sigma_t=-\sigma_b$. To explain the underlying physics behind these results, we first observe the space charge density along the nanochannel axis $Y=0$ in figure~\ref{fig_result_4}f. Figure~\ref{fig_result_4}f depicts that the space charge density vanishes along the channel axis when $\sigma_t=-\sigma_b$, attributed primarily to the zero electric potential (figure~\ref{fig_result_4}b) along the channel axis $Y=0$. Is implies that the axial electric field (figure~\ref{fig_result_4}c) and corresponding axial pressure gradient (figure~\ref{fig_result_4}d) also vanish along $Y=0$. However, pertaining to the overlapping EDL scenarios, the electric potential does not vanish along the nanochannel axis for $\sigma_t=\sigma_b$. This phenomenon helps the axial velocity attain a maximum value along the channel axis; thus, the velocity profile also tends to become parabolic. As a result, the magnitude of COF axial velocity for overlapping EDL cases becomes higher for the charge densities specified at both walls having the same sign.

\subsection{Numerical results beyond low surface charge consideration}
Generally, the magnitude of axial COF velocity is comparatively smaller than the magnitude of axial EOF velocity, typically obtained within a nanochannel for a certain window of surface charge density. In this section, we will discuss several facets leading toward obtaining COF with a sufficiently higher magnitude of axial velocity, comparable with the EOF velocity typically obtained in a nanochannel. It is worth mentioning that the results presented in this subsection correspond to beyond Debye-H\"uckel approximation as the magnitude of surface charge densities considered is higher.

\begin{figure}
  \centering
  \subfigure{\includegraphics[width=2.25in]{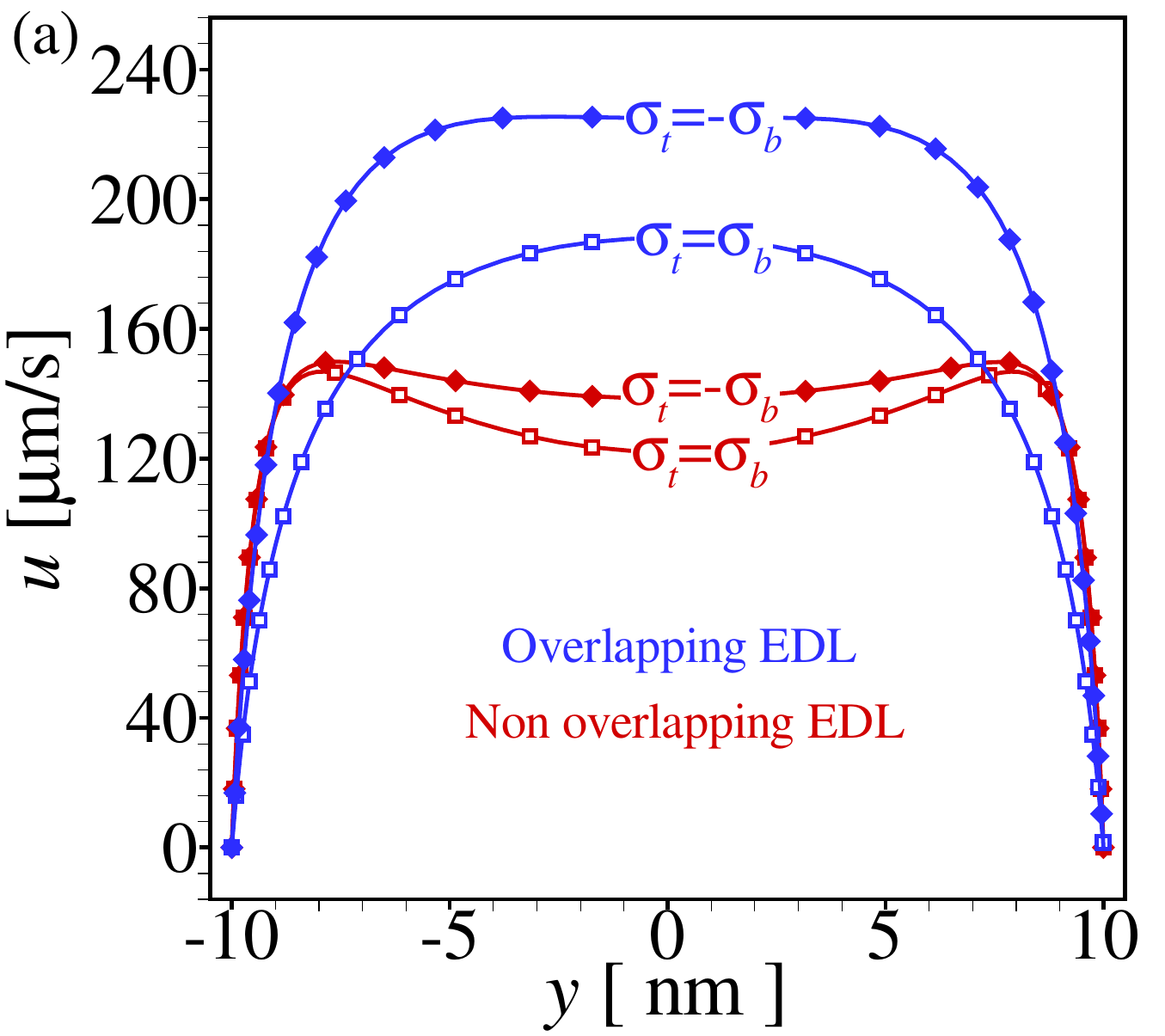}}
  \subfigure{\includegraphics[width=2.25in]{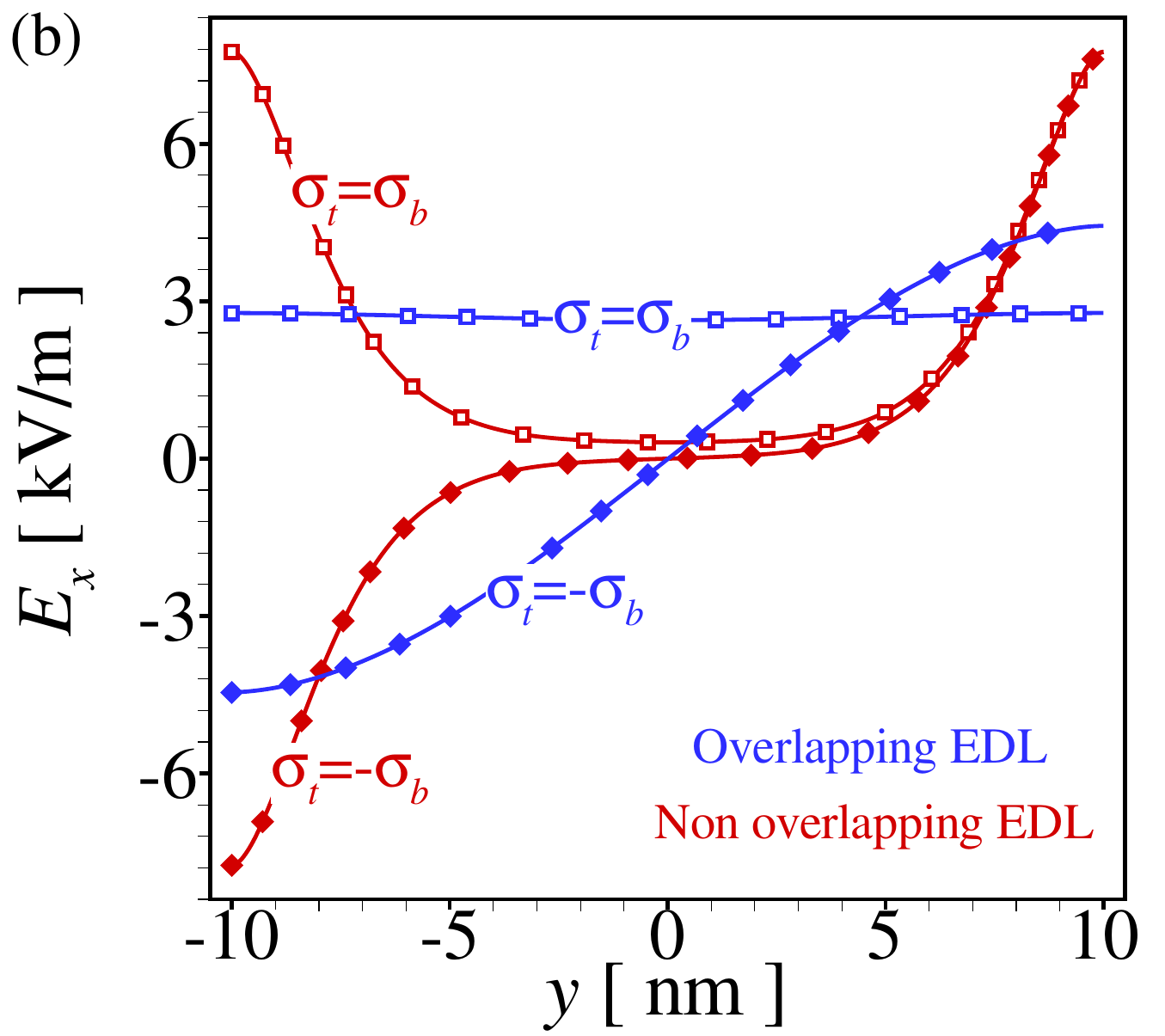}} 
  \caption{(a) Axial COF velocity and corresponding (b) axial electric field across the nanochannel cross-section for $\sigma_t=-60$ mC/m$^2$ and $\sigma_t=\pm60$ mC/m$^2$ at overlapping EDL ($c_L=2$ mM, $c_R=1$ mM) and nonoverlapping EDL ($c_L=100$ mM, $c_R=50$ mM).} 
  \label{fig_result_5}
\end{figure}

Figure~\ref{fig_result_5}a,b, respectively, plots the COF axial velocity and the corresponding axial electric field for $\sigma_t=-60$ mC/m$^2$ and $\sigma_t=\pm60$ mC/m$^2$. The variations presented in Fig.~\ref{fig_result_5}a,b are obtained for both the cases of overlapping ($c_L=2$ mM, $c_R=1$ mM) and nonoverlapping EDL ($c_L=100$ mM, $c_R=50$ mM). Looking at the results obtained under the Debye-H\"uckel limit, as presented in figures~\ref{fig_result_1}a, \ref{fig_result_2}a, \ref{fig_result_4}a, we observe that the magnitude of axial COF velocity is higher for overlapping EDL than nonoverlapping EDL. Also, as witnessed in Fig.~\ref{fig_result_5}, this behavior holds true even for higher surface charge density beyond the Debye-H\"uckel limit. However, unlike the variations of flow velocity pertaining to the Debye-H\"uckel cases that we discussed in previous subsections, the axial COF velocity is seen to be higher when the sign of both wall charge densities are opposite, i.e., $\sigma_t=-\sigma_b$ ( cf. figure~\ref{fig_result_5}a ). In particular, this aspect is deemed significant for overlapping EDL cases as witnessed in figure~\ref{fig_result_5}a. In an effort to delve deep into the physical reasoning behind this aspect, we plot the corresponding axial electric field in figure~\ref{fig_result_5}b. For overlapping EDL, the axial electric field remains almost the same throughout the cross-section when $\sigma_b=\sigma_t$. Also, the axial electric field changes the sign on both sides of the nanochannel when $\sigma_t=-\sigma_b$. As seen from the variation demonstrated in Fig.~\ref{fig_result_5}b for overlapping EDL cases, the magnitude of the axial electric field in the vicinity of the walls attains a higher value when the signs of wall charge densities are opposite. This observation underlines that a substantially higher axial COF velocity is obtained for the oppositely signed wall charge densities. On the other hand, this behavior of the axial electric field does not hold for the nonoverlapping EDL scenarios. Thus, a slight discrepancy in the axial COF velocities between $\sigma_b=\pm\sigma_t$ appears only near the channel axis $Y=0$. Moreover, the velocity profiles for these nonoverlapping cases deviate from the plug-like shape, especially for $\sigma_t=\sigma_b$. We attribute this observation as follows. Pertaining to the case of $\sigma_b=\sigma_t$, the axial electric field, which solely generates from the EDL potential, does not vanish, and the space charge density also does not disappear near the channel axis $Y=0$. The conjugated effect of these two factors results in a deviation of the flow velocity from the plug-like shape. On the other hand, along the channel axis $Y=0$ when $\sigma_b=-\sigma_t$, the non-existing space charge density compelled the axial electric field to vanish therein. Thus, the deviation from achieving the plug-like velocity profile is reduced for surface charge densities having the same magnitude but unequal sign.

\section{Conclusion}
Our analysis thoroughly investigates the characteristics of the chemiosmotic flow (COF) within a reservoir-connected nanochannel having uniform surface charge density on the walls. We derive an analytical solution for the axial COF velocity under the framework of the lubrication approximation. In the limit of the Debye-H\"uckel approximation, we obtain analytical solutions for both cases of overlapping or non-overlapping electric double layers. We develop a 2D numerical model consistent with the finite element framework of COMSOL multiphysics to describe the COF within a finite-length nanochannel connected with identical reservoirs on both ends. The modeling framework employs a coupled set of non-linear Poisson-Nernst-Planck-Navier-Stokes equations to solve the underlying transport features. An optimum set of structured grid distributions is finalized by performing a proper grid-independence test, and selected grids are used for all the cases simulated in this analysis. Both the numerical and analytical models are further validated by comparing results obtained from each framework in the Debye-H\"uckel limit. Results show an excellent agreement between the analytical and numerical solutions in the limit of the Debye-H\"uckel approximation. Moreover, we analytically derive and establish an effective velocity scale of COF for non-overlapping EDLs. Also, we demonstrate the numerical results of COF velocity, obtained beyond the Debye-H\"uckel limit. 

Below we summarize the significant characteristics of the chemiosmotic flow, analyzed in the present study, as follows:
\textbf{(1)} The magnitude of the COF velocity depends upon the magnitude of the concentration gradient and charge densities of the channel walls. 
\textbf{(2)} the magnitude of the axial COF velocity is inversely proportional to the bulk concentration.
\textbf{(3)} The direction of the axial COF velocity only depends upon the direction of the concentration gradient. Specifically, the COF velocity direction is always from the higher to the lower concentration. 
\textbf{(4)} Unlike EOF, the direction of the axial COF velocity does not depend upon the sign of surface charge density of the channel walls. 
\textbf{(5)} Pertaining to sufficiently thinner EDL, i.e., non-overlapping case and under the purview of Debye-H\"uckel approximation, the axial COF velocity obtained for both the same signed and oppositely signed wall charge densities are identical. 
\textbf{(6)} For overlapping EDLs and in the Debye-H\"uckel limit, the velocity obtained for symmetric wall charge densities ($\sigma_b=\sigma_t$: same in magnitude and sign) is higher than the velocity obtained for asymmetric wall charge densities ($\sigma_b=-\sigma_t$: same in magnitude but opposite in sign). 
\textbf{(7)} Beyond the Debye-H\"uckel limit, the velocity obtained for $\sigma_b=\sigma_t$ is lower than the velocity obtained for $\sigma_b=-\sigma_t$.

\section*{Acknowledgement} 
P.K.M gratefully acknowledges the financial support provided by the SERB (DST), India, through Project No. MTR/2020/000034.

\section*{Declaration of interests}
The author reports no conflict of interest.

\bibliographystyle{jfm}
% Note the spaces between the initials
\bibliography{jfm-instructions}

\end{document}